\documentclass[9pt,twocolumn,twoside]{osajnl}

\journal{josaa} 

\setboolean{shortarticle}{false}

\usepackage{graphicx}
\usepackage{wrapfig}
\usepackage{subfigure}
\usepackage{float}
\usepackage{mathtools}
\usepackage[capitalize]{cleveref}
\usepackage[color]{changebar}
\cbcolor{red}

\title{Spatio-temporal statistics of the turbulent piston-removed phase and Zernike coefficients for two distinct beams}

\author[1,*]{C{\'e}dric Plantet}
\author[1]{Giulia Carl{\`a}}
\author[1]{Guido Agapito}
\author[1]{Lorenzo Busoni}

\affil[1]{INAF, Osservatorio Astrofisico di Arcetri, Largo Enrico Fermi 5, 50125 Firenze, Italy}

\affil[*]{cedric.plantet@inaf.it}

\begin{abstract}
In the context of adaptive optics for astronomy, one can rely on the statistics of the turbulent phase to assess a part of the system's performance. Temporal statistics with one source and spatial statistics with two sources are well-known and are widely used for classical adaptive optics systems. A more general framework, including both spatial and temporal statistics, can be useful 
for 
the analysis of the existing systems and to support the design of the future ones.
In this paper, we propose an expression of the 
temporal cross power spectral densities 
of the turbulent phases in two distinct beams, that is from two different sources to two different apertures. We either consider the phase as it is, without piston, or as its decomposition on Zernike modes. The general formulas allow to cover a wide variety of configurations, from single-aperture to interferometric telescopes equipped with adaptive optics, 
with the possibility to consider apertures of different sizes and/or sources at a finite distance. 
The presented approach should lead to similar results with respect to existing methods in the Fourier domain, but it is focused on temporal frequencies rather than spatial ones, which might be convenient for some aspects such as control optimization. 
To illustrate this framework with a simple application, we demonstrate that the wavefront residual due to the anisoplanatism error in a single-conjugated adaptive optics system is overestimated when it is computed from covariances without taking into account the temporal filtering of the adaptive optics loop. We also show this overestimation in the case of a small-baseline interferometer, for which the two beams are significantly correlated. 
\end{abstract}

\setboolean{displaycopyright}{true}

\begin{document}
\maketitle
\section{Introduction}
In the context of adaptive optics (AO) for astronomy, one can rely on the statistics of the turbulent phase to assess and optimize a part of the system's performance. Many studies of the turbulence statistics have been done considering a wavefront decomposition on the Zernike modes \cite{noll,sasiela,conan,rconan,navarro,roddier,nroddier,gendron,negro,molodij,hu,takato,whiteley}. They indeed represent an orthogonal basis on a circular aperture \cite{born}, that is the most common aperture shape in optical systems. These studies mostly focus on the temporal statistics of the turbulence seen from one source to one aperture \cite{conan,roddier,hogge} or the spatial covariance from one or two sources to one or two apertures \cite{noll,sasiela,rconan,conan,takato, whiteley}. Though, the knowledge of both temporal and spatial statistics in a general framework can be useful for the development of new analytical expressions 
to estimate 
the adaptive optics performance. These tools can help 
in 
the analysis of existing Single-Conjugated Adaptive Optics (SCAO) or Wide-Field AO (WFAO) systems \cite{pinna2016soul,petit2016saxo,neichel2010gemini,stuik2006galacsi} and future systems that are going to equip the 
next generation 
of telescopes \cite{diolaiti,neichel,herriot,hinz,mcdermid,conan2012giant}. Indeed, the classical approach for the analysis of the performance of an AO system is to decompose the overall residual in several sources of errors (temporal, anisoplanatism, noise, aliasing, fitting\dots), considering them uncorrelated \cite{madec,clenet,gendron2014novel,sandler,neichel2009tomographic,rigaut1998analytical}. In that case, most of the error computations do not take into account the temporal filtering of the AO loop, while alternative approaches \cite{ellerbroek2005linear,jolissaint,clare2006adaptive,correia2017modeling} apply the AO control in the whole performance analysis, highlighting for example the correlation between the temporal and the anisoplanatism errors. 
These methods to evaluate the AO performance often rely on an analysis in the spatial frequency domain \cite{ellerbroek2005linear,jolissaint,neichel2009tomographic,rigaut1998analytical,clenet,gendron2014novel}. 

In this paper, we use the geometry presented by Whiteley \textit{et al.} \cite{whiteley}, 
who expressed the spatial and temporal covariance of Zernike modes for two different sources and two different apertures, to derive temporal Cross Power Spectral Densities (CPSDs) of the piston-removed phase seen on two distinct beams. 
We slightly modify the formalism in order to include the case of sources at infinity. We use our calculations to develop an expression of the CPSDs of two Zernike coefficients as well, starting from the one that has been introduced by Whiteley \cite{whiteley}. 
To our knowledge, this framework is the only one that offers the possibility to directly take into account the following aspects altogether in a single formula, while allowing the application of a time filtering: distinct apertures of different size, distinct sources at finite or infinite distance, direction of the wind. It should provide similar results with respect to spatial-frequency-based approaches, given the same assumptions, but it is focused on the temporal frequencies. It thus allows a simpler exploration of a different dimension of the problem, which might be convenient for control optimization for example. 
We then present a case study that makes use of the CPSDs to derive the anisoplanatism error for a SCAO system and for an interferometer such as the Large Binocular Telescope Interferometer (LBTI) \cite{hinz_lbti16}. We show that a 
simple 
computation from covariances would overestimate the anisoplanatism error with respect to a more precise computation that takes into account the temporal filtering of the AO loop. 

In section \ref{secgeom}, we present the aperture-source geometry used throughout the paper. In section \ref{secphasecov}, we give the expression of the inter-aperture spatial covariance of the piston-removed phase, that we then use to compute the corresponding spatio-temporal CPSDs (section \ref{secphasepsd}). In section \ref{secpsd}, we use the formalism introduced in section \ref{secphasepsd} to extend the expression of the Zernike coefficients CPSDs. Finally, in section \ref{numver}, we present the case study where we consider a SCAO correction on either a single-aperture or a two-aperture interferometric telescope.

\section{Geometry}
\label{secgeom}

\begin{figure}[htbp]
\centering
\includegraphics[width=\linewidth]{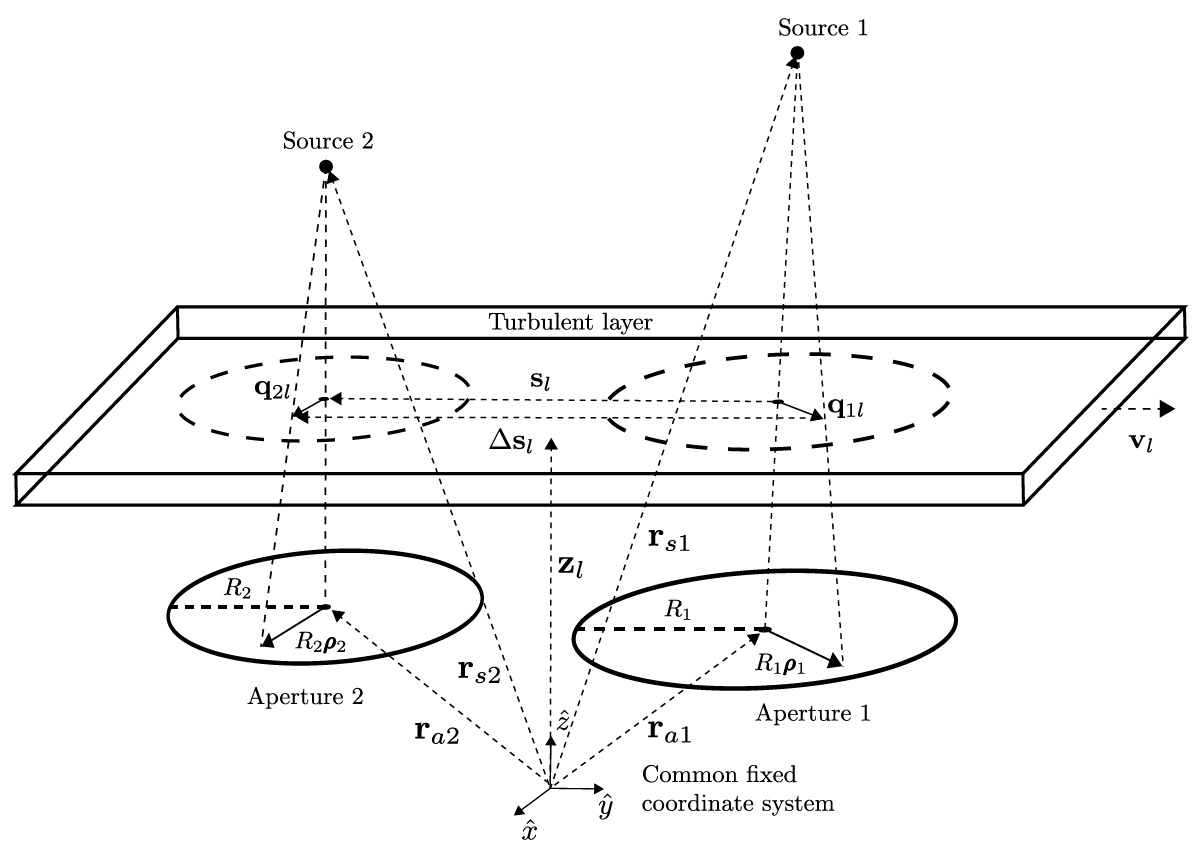}
\caption{Geometry used to compute the spatio-temporal CPSDs.}
\label{figgeom}
\end{figure}

The aperture-source geometry we consider here is the one introduced by Whiteley \textit{et al.} \cite{whiteley}, that is reproduced in Fig.~\ref{figgeom}. We have two apertures of radii $R_1$ and $R_2$ (located by the vectors $\mathbf{r}_{a1}$ and $\mathbf{r}_{a2})$ observing two different sources (located by the vectors $\mathbf{r}_{s1}$ and $\mathbf{r}_{s2}$) through a turbulent layer at altitude $\mathbf{z}_l$. A ray coming from the first (respectively the second) source and arriving at a point located by the vector $R_1 \pmb{\rho}_1$ (resp. $R_2 \pmb{\rho}_2$) with respect to the first (resp. the second) aperture center will pass by the point located by $\mathbf{q}_{1l}$ (resp. $\mathbf{q}_{2l}$) in the aperture footprint in the turbulent layer. The projected vectors $\mathbf{q}_{1l}$ and $\mathbf{q}_{2l}$ are expressed as follows:
\begin{equation}
\label{eqrho2q1}
\mathbf{q}_{1l} = (1-A_{1l}) R_1 \pmb{\rho}_1
\end{equation}
\begin{equation}
\label{eqrho2q2}
\mathbf{q}_{2l} = (1-A_{2l}) R_2 \pmb{\rho}_2
\end{equation}
where $A_{1l}$ and $A_{2l}$ are the layer scaling factors:
\begin{equation}
\label{eqA1}
A_{1l} = \frac{z_l-\mathbf{r}_{a1}\cdot \hat{z}}{(\mathbf{r}_{s1}-\mathbf{r}_{a1})\cdot \hat{z}}
\end{equation}
\begin{equation}
\label{eqA2}
A_{2l} = \frac{z_l-\mathbf{r}_{a2}\cdot \hat{z}}{(\mathbf{r}_{s2}-\mathbf{r}_{a2})\cdot \hat{z}}
\end{equation}
If $\mathbf{r}_{a1} = \mathbf{r}_{a2} = \mathbf{0}$, then \eqref{eqA1} and \eqref{eqA2} simplify into:
\begin{equation}
A_{1l} = \frac{z_l}{z_1}
\end{equation}
\begin{equation}
A_{2l} = \frac{z_l}{z_2}
\end{equation}
with $z_1$ and $z_2$ the sources' altitudes.

In the following, we will need to express the vector joining two points of the apertures' footprints in the turbulent layer:
 \begin{equation}
\Delta \mathbf{s}_l = \mathbf{q}_{2l}-\mathbf{q}_{1l}+\mathbf{s}_l
\end{equation}
with $\mathbf{s}_l$ the vector joining the centers of the footprints:
\begin{equation}
\mathbf{s}_l = \mathbf{r}_{a2}-\mathbf{r}_{a1}+\frac{z_l-\mathbf{r}_{a2}\cdot \hat{z}}{\mathbf{r}_{s2}'\cdot \hat{z}}\mathbf{r}_{s2}'-\frac{z_l-\mathbf{r}_{a1}\cdot \hat{z}}{\mathbf{r}_{s1}'\cdot \hat{z}}\mathbf{r}_{s1}'
\end{equation}
with $\mathbf{r}_{s1}' = \frac{\mathbf{r}_{s1}-\mathbf{r}_{a1}}{|\mathbf{r}_{s1}-\mathbf{r}_{a1}|}$ (resp. $\mathbf{r}_{s2}' = \frac{\mathbf{r}_{s2}-\mathbf{r}_{a2}}{|\mathbf{r}_{s2}-\mathbf{r}_{a2}|}$) the unitary vector from the center of aperture 1 (resp. 2) to source 1 (resp. 2). This formula is slightly different from the one introduced by Whiteley \textit{et al.} \cite{whiteley} to take into account the case of sources at infinity. 
In the following, we derive the covariance of the piston-removed phase, that had not been considered by Whiteley \textit{et al.}, in order to then find the temporal CPSD of this same quantity. Indeed,  considering the full phase instead of its decomposition on wavefront modes can be convenient as it allows a much faster estimation of the full wavefront error. 

\section{Inter-aperture covariance of the piston-removed phase}
\label{secphasecov}
In this section, we consider the whole phases $\phi_1$ and $\phi_2$ in the respective apertures 1 and 2, without any decomposition on wavefront modes. We define their piston-filtered covariance as:
\begin{equation}
\label{eqphasecov1}
\begin{split}
C_{\phi_1,\phi_2} = & \text{E}\Bigg\{\int d\pmb{\rho} \left[\phi_1(R_1\pmb{\rho})-\int d\pmb{\rho}' \phi_1(R_1\pmb{\rho}') P(\pmb{\rho}') \right] \\
& \times \left[\phi_2(R_2\pmb{\rho})-\int d\pmb{\rho}'' \phi_2(R_2\pmb{\rho}'') P(\pmb{\rho}'') \right] P(\pmb{\rho}) \Bigg\}
\end{split}
\end{equation}
where $\pmb{\rho} = \pmb{\rho}_1 = \pmb{\rho}_2$ if one refers to Fig.~\ref{figgeom}. $\text{E}\{\ \}$ is the mathematical expectation and $P(\pmb{\rho})$ is the aperture weighting function:
\begin{equation}
P(\pmb{\rho}) = 
\begin{dcases}
\frac{1}{\pi} & \text{if} \ |\pmb{\rho}| \leq 1 \\
0 & \text{otherwise}
\end{dcases}
\end{equation}
We develop \eqref{eqphasecov1}:
\begin{equation}
\label{eqphasecov2}
\begin{split}
C_{\phi_1,\phi_2} = & \int d\pmb{\rho} \text{E}\{\phi_1(R_1\pmb{\rho})\phi_2(R_2\pmb{\rho})\}  P(\pmb{\rho}) \\
& - \int d\pmb{\rho} \int d\pmb{\rho}' \text{E}\{\phi_1(R_1\pmb{\rho}')\phi_2(R_2\pmb{\rho})\} P(\pmb{\rho}') P(\pmb{\rho}) \\
& - \int d\pmb{\rho} \int d\pmb{\rho}'' \text{E}\{\phi_1(R_1\pmb{\rho})\phi_2(R_2\pmb{\rho}'')\} P(\pmb{\rho}'') P(\pmb{\rho})  \\
 & + \int d\pmb{\rho}\  P(\pmb{\rho}) \int d\pmb{\rho}' \int d\pmb{\rho}'' \text{E}\{\phi_1(R_1\pmb{\rho}')\phi_2(R_2\pmb{\rho}'')\} \\
& \times P(\pmb{\rho}') P(\pmb{\rho}'') 
\end{split}
\end{equation}
We notice that the second, the third and the last integral are equivalent, given that $\int d\pmb{\rho}\  P(\pmb{\rho})=1$. Besides, $C_{\phi_1,\phi_2}$ depends on the phase cross-correlation that, when considering independent turbulent layers, can be expressed as:
\begin{equation}
\text{E}\left\{\phi_1(R_1 \pmb{\rho}_1) \phi_2(R_2 \pmb{\rho}_2)\right\} = \sum_l B_{\phi_l}(\mathbf{q}_{1l},\mathbf{q}_{2l})
\end{equation}
with $\mathbf{q}_{1l}$ and $\mathbf{q}_{2l}$ as defined by \eqref{eqrho2q1} and \eqref{eqrho2q2}. 
Assuming that the turbulent phase is spatially stationary, the cross-correlation $B_{\phi_l}(\mathbf{q}_{1l},\mathbf{q}_{2l})$  only depends on the vector separating the two considered points:
\begin{equation}
B_{\phi_l}(\mathbf{q}_{1l},\mathbf{q}_{2l}) = B_{\phi_l}(\Delta \mathbf{s}_l) = B_{\phi_l}(\mathbf{q}_{2l}-\mathbf{q}_{1l}+\mathbf{s}_l)
\end{equation}
We then have:
\begin{equation}
\begin{split}
C_{\phi_1,\phi_2} = & \sum_l C_{\phi_1,\phi_2,l} = \sum_l \Bigg[ \frac{1}{R_1 (1-A_{1l})} \int d\mathbf{q}_{1l} \\
& \times B_{\phi_l}(\mathbf{q}_{2l}-\mathbf{q}_{1l}+\mathbf{s}_l) P\left(\frac{\mathbf{q}_{1l}}{R_1 (1-A_{1l})}\right) \\
& - \frac{1}{R_1 R_2 (1-A_{1l})(1-A_{2l})} \int d\mathbf{q}_{1l} \int d\mathbf{q}_{2l}' \\
& \times B_{\phi_l}(\mathbf{q}_{2l}'-\mathbf{q}_{1l}+\mathbf{s}_l) P\left(\frac{\mathbf{q}_{2l}'}{R_2 (1-A_{2l})}\right) \\
& \times P\left(\frac{\mathbf{q}_{1l}}{R_1 (1-A_{1l})}\right)\Bigg]
\end{split}
\end{equation}
with $\mathbf{q}_{2l}' = (1-A_{2l}) R_2 \pmb{\rho}''$. Since $\pmb{\rho} = \pmb{\rho}_1 = \pmb{\rho}_2$, we must have:
\begin{equation}
\label{eqK}
\mathbf{q}_{2l} = (1-A_{2l}) R_2 \pmb{\rho} = \frac{(1-A_{2l}) R_2}{(1-A_{1l}) R_1} \mathbf{q}_{1l} = K \mathbf{q}_{1l}
\end{equation}
We evaluate the first integral using the variable change $\mathbf{q} = (K-1) \mathbf{q}_{1l}$:
\begin{equation}
\begin{split}
T_1 & = \frac{1}{R_1 (1-A_{1l})} \int d\mathbf{q}_{1l} B_{\phi_l}(\mathbf{q}_{2l}-\mathbf{q}_{1l}+\mathbf{s}_l)  P\left(\frac{\mathbf{q}_{1l}}{R_1 (1-A_{1l})}\right) \\
& = \frac{1}{R_1 (1-A_{1l}) (K-1)} \int d\mathbf{q} B_{\phi_l}(\mathbf{q}+\mathbf{s}_l) \\
& \times P\left(\frac{\mathbf{q}}{R_1 (1-A_{1l})(K-1)}\right)
\end{split}
\end{equation}
Using Parseval's theorem, we can write $T_1$ as:
\begin{equation}
\begin{split}
T_1 = & \frac{1}{R_1 (1-A_{1l}) (K-1)} \int d\mathbf{f}\ \text{FT}\{B_{\phi_l}(\mathbf{q}+\mathbf{s}_l)\}^* \\
& \times \text{FT}\left\{P\left(\frac{\mathbf{q}}{R_1 (1-A_{1l})(K-1)}\right)\right\}
\end{split}
\end{equation}
where $\text{FT}\{\ \}$ is the Fourier transform (from $\mathbf{q}$ to $\mathbf{f}$ in this case). 
The Fourier transform of the phase correlation is given by Wiener-Khinchin's theorem:
\begin{equation}
\text{FT}\left\{B_{\phi_l}(\mathbf{q}+\mathbf{s}_l)\right\} = W_{\phi_l}(\mathbf{f}) \exp[2i\pi \mathbf{f} \cdot \mathbf{s}_l]
\end{equation}
where $W_{\phi_l}$ is the spatial power spectrum of the turbulent phase in the layer $l$, often assumed to follow Von Karman's model \cite{rconan}:
\begin{equation}
W_{\phi_l}(f) = \left[\frac{24}{5} \Gamma\left(\frac{6}{5}\right)\right]^{\frac{5}{6}} \frac{\Gamma\left(\frac{11}{6}\right)^2}{2 \pi^{\frac{11}{3}}} r_0^{-\frac{5}{3}} \left(f^2+\frac{1}{L_0^2}\right)^{-\frac{11}{6}}
\end{equation}
with $\Gamma(\ )$ the gamma function, $r_0$ the Fried parameter and $L_0$ the outer scale. Hence, we finally have:
\begin{equation}
T_1 = \int d\mathbf{f}\ W_{\phi_l}(\mathbf{f}) \exp[-2i\pi \mathbf{f} \cdot \mathbf{s}_l] \frac{J_1 \left(2\pi R_1 (1-A_{1l}) (K-1) f \right)}{\pi R_1 (1-A_{1l}) (K-1) f}
\end{equation}
with $J_1$ the Bessel function of the first kind and order 1. The second integral to evaluate is:
\begin{equation}
\begin{split}
 T_2 = & \frac{1}{R_1 R_2 (1-A_{1l})(1-A_{2l})} \int d\mathbf{q}_{1l} \int d\mathbf{q}_{2l}' B_{\phi_l}(\mathbf{q}_{2l}'-\mathbf{q}_{1l}+\mathbf{s}_l) \\
&  \times  P\left(\frac{\mathbf{q}_{2l}'}{R_2 (1-A_{2l})}\right) P\left(\frac{\mathbf{q}_{1l}}{R_1 (1-A_{1l})}\right)
\end{split}
\end{equation}
Again, using Parseval's and Wiener-Khinchin's theorems with a Fourier transform on $\mathbf{q}_{2l}'$, we find:
\begin{equation}
\begin{split}
T_2 = & \frac{1}{R_1 (1-A_{1l})} \int d\mathbf{q}_{1l} \int d\mathbf{f}\ W_{\phi_l}(\mathbf{f}) \exp[2i\pi \mathbf{f} \cdot (\mathbf{q}_{1l}-\mathbf{s}_l)] \\
& \times \frac{J_1 \left(2\pi R_2 (1-A_{2l}) f \right)}{\pi R_2 (1-A_{2l}) f} P\left(\frac{\mathbf{q}_{1l}}{R_1 (1-A_{1l})}\right)
\end{split}
\end{equation}
When re-ordering the integrals, one finds a Fourier transform in $\mathbf{q}_{1l}$, leading to:
\begin{equation}
\begin{split}
T_2 = & \int d\mathbf{f} W_{\phi_l}(\mathbf{f}) \exp[-2i\pi \mathbf{f} \cdot \mathbf{s}_l] \frac{J_1 \left(2\pi R_1 (1-A_{1l}) f \right)}{\pi R_1 (1-A_{1l}) f} \\
& \times \frac{J_1 \left(2\pi R_2 (1-A_{2l}) f \right)}{\pi R_2 (1-A_{2l}) f}
\end{split}
\end{equation}
The piston-filtered covariance for the layer $l$ is then:
\begin{equation}
\label{eqphasecov}
\begin{split}
C_{\phi_1,\phi_2,l} = & T_1 - T_2 = \int d\mathbf{f}\ W_{\phi_l}(\mathbf{f}) \exp[-2i\pi \mathbf{f} \cdot \mathbf{s}_l] \\
& \times \Bigg[ \frac{J_1 \left(2\pi R_1 (1-A_{1l}) (K-1) f \right)}{\pi R_1 (1-A_{1l}) (K-1) f} \\
&  - \frac{J_1 \left(2\pi R_1 (1-A_{1l}) f \right)}{\pi R_1 (1-A_{1l}) f} \frac{J_1 \left(2\pi R_2 (1-A_{2l}) f \right)}{\pi R_2 (1-A_{2l}) f} \Bigg]
\end{split}
\end{equation}
For a single aperture and sources at infinity, the last term of the integral becomes the classical filter function for piston removal $1-\left[\frac{J_1 \left(2\pi R f \right)}{\pi R f} \right]^2$ \cite{sasiela} (we remind that $\frac{J_1 \left(0 \right)}{0}=1$). 
When integrating over the angle, we find:
\begin{equation}
\begin{split}
C_{\phi_1,\phi_2,l} = & 2\pi \int_0^{\infty} f df\ W_{\phi_l}(f) J_0 \left(2\pi f s_l \right)\\
& \times \Bigg[ \frac{J_1 \left(2\pi R_1 (1-A_{1l}) (K-1) f \right)}{\pi R_1 (1-A_{1l}) (K-1) f} \\
&  - \frac{J_1 \left(2\pi R_1 (1-A_{1l}) f \right)}{\pi R_1 (1-A_{1l}) f} \frac{J_1 \left(2\pi R_2 (1-A_{2l}) f \right)}{\pi R_2 (1-A_{2l}) f} \Bigg]
\end{split}
\end{equation}
\section{Spatio-temporal cross power spectrum of the piston-removed phase}
\label{secphasepsd}
\begin{figure}[htbp]
\centering
\includegraphics[width=.7\linewidth]{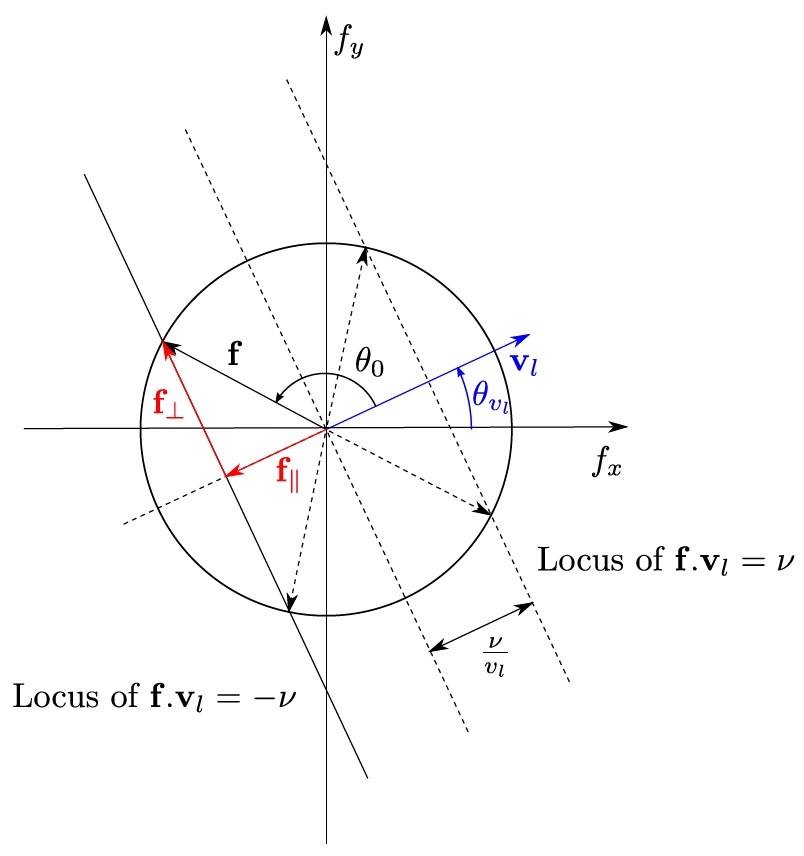}
\caption{Vectors and angles defined in the CPSD computation. Two lines represent the loci of $\mathbf{f} \cdot  \mathbf{v}_l = \nu$ (dashed line on right-hand side) and $\mathbf{f} \cdot  \mathbf{v}_l = - \nu$ (solid line), considering a positive $\nu$. For a given frequency, the Dirac defined in \eqref{eqdirac} reduces the CPSD expression to the sum of 2 points, either on the left or the right side of the figure, that are indicated by $\mathbf{f}$ and the dashed vectors. In this figure, we have $\mathbf{f} = \mathbf{f}_1$ (see \eqref{eqphasepsd3}).}
\label{figgeom2}
\end{figure}

We now consider that we observe the first source at a time $t=0$ and the second source at $t=\tau$. Here, we assume a motion of the turbulent layer following Taylor's frozen flow hypothesis along the wind vector $\mathbf{v}_l$ (Fig.~\ref{figgeom}), while the sources and apertures remain fixed. We can then define the effective footprint separation as a function of $\tau$ in the layer $l$:
\begin{equation}
\mathbf{s}'_l(\tau) = \mathbf{s}_l-\mathbf{v}_l \tau
\end{equation}
The spatio-temporal cross-correlation is then (from \eqref{eqphasecov}):
\begin{equation}
\begin{split}
R_{\phi_1,\phi_2,l} (\tau) = & \int d\mathbf{f}\ W_{\phi_l}(\mathbf{f}) \exp[-2i\pi \mathbf{f} \cdot \mathbf{s}'_l(\tau)] \\
& \times \Bigg[ \frac{J_1 \left(2\pi R_1 (1-A_{1l}) (K-1) f \right)}{\pi R_1 (1-A_{1l}) (K-1) f} \\
&  - \frac{J_1 \left(2\pi R_1 (1-A_{1l}) f \right)}{\pi R_1 (1-A_{1l}) f} \frac{J_1 \left(2\pi R_2 (1-A_{2l}) f \right)}{\pi R_2 (1-A_{2l}) f} \Bigg]
\end{split}
\end{equation}
If $\tau = 0$, one retrieves $C_{\phi_1,\phi_2,l}$. The spatio-temporal CPSD of the piston-removed phase is the Fourier transform of its cross-correlation:
\begin{equation}
S_{\phi_1,\phi_2,l} (\nu) = \int d\tau\ R_{\phi_1,\phi_2,l}(\tau) \exp[2i\pi \nu \tau]
\end{equation}
\begin{equation}
\begin{split}
S_{\phi_1,\phi_2,l} (\nu) = & \int d\tau\ \int d\mathbf{f}\ W_{\phi_l}(\mathbf{f}) \exp[-2i\pi \mathbf{f} \cdot \mathbf{s}'_l(\tau)] \\
& \times \Bigg[ \frac{J_1 \left(2\pi R_1 (1-A_{1l}) (K-1) f \right)}{\pi R_1 (1-A_{1l}) (K-1) f} \\
&  - \frac{J_1 \left(2\pi R_1 (1-A_{1l}) f \right)}{\pi R_1 (1-A_{1l}) f} \frac{J_1 \left(2\pi R_2 (1-A_{2l}) f \right)}{\pi R_2 (1-A_{2l}) f} \Bigg] \\
& \times \exp[2i\pi \nu \tau]
\end{split}
\end{equation}
where $\nu$ is the temporal frequency. We replace $\mathbf{s}'_l (\tau)$ with $\mathbf{s}_l-\mathbf{v}_l \tau$:
\begin{equation}
\label{eqphasepsd1}
\begin{split}
S_{\phi_1,\phi_2,l} (\nu) = & \int d\mathbf{f}\ W_{\phi_l}(\mathbf{f}) \exp[-2i\pi \mathbf{f} \cdot  \mathbf{s}_l] \\
& \times \Bigg[ \frac{J_1 \left(2\pi R_1 (1-A_{1l}) (K-1) f \right)}{\pi R_1 (1-A_{1l}) (K-1) f} \\
&  - \frac{J_1 \left(2\pi R_1 (1-A_{1l}) f \right)}{\pi R_1 (1-A_{1l}) f} \frac{J_1 \left(2\pi R_2 (1-A_{2l}) f \right)}{\pi R_2 (1-A_{2l}) f} \Bigg] \\
& \times \int d\tau\  \exp[2i\pi (\nu+\mathbf{f} \cdot  \mathbf{v}_l) \tau]
\end{split}
\end{equation}
We now consider the components of $\mathbf{f}$, $\mathbf{f}_{\perp}$ and $\mathbf{f}_{\parallel}$, so that $\mathbf{f}_{\perp}$ is orthogonal to $\mathbf{v}_l$ and $\mathbf{f}_{\parallel}$ is parallel to $\mathbf{v}_l$ (see Fig.~\ref{figgeom2}). We also define the unitary vector along the wind direction $\hat{u} = \frac{\mathbf{v}_l}{v_l}$. \eqref{eqphasepsd1} can be written:
\begin{equation}
\label{eqphasepsd2}
\begin{split}
S_{\phi_1,\phi_2,l} (\nu) = & \int d\mathbf{f}_{\perp} \int d\mathbf{f}_{\parallel} \ W_{\phi_l}(\mathbf{f}_{\parallel},\mathbf{f}_{\perp}) \exp[-2i\pi \mathbf{f} \cdot  \mathbf{s}_l] \\
& \times \Bigg[ \frac{J_1 \left(2\pi R_1 (1-A_{1l}) (K-1) f \right)}{\pi R_1 (1-A_{1l}) (K-1) f} \\
&  - \frac{J_1 \left(2\pi R_1 (1-A_{1l}) f \right)}{\pi R_1 (1-A_{1l}) f} \frac{J_1 \left(2\pi R_2 (1-A_{2l}) f \right)}{\pi R_2 (1-A_{2l}) f} \Bigg] \\
& \times \int d\tau\ \exp[2i\pi (\frac{\nu}{v_l}+ \mathbf{f}_{\parallel}\cdot \hat{u}) v_l \tau]
\end{split}
\end{equation}
The last integral is a Dirac function. If we consider that $f_{\parallel}$ is positive when $\mathbf{f}_{\parallel}$ is pointing towards the same direction as $\mathbf{v}_l$ and negative otherwise, then:
\begin{equation}
\begin{split}
\label{eqdirac}
& \int d\tau\ \exp[2i\pi (\frac{\nu}{v_l}+\mathbf{f}_{\parallel}\cdot \hat{u}) v_l \tau] = \frac{1}{v_l} \int dt\ \exp[2i\pi (\frac{\nu}{v_l}+f_{\parallel}) t] \\
& = \frac{1}{v_l} \delta\left(f_{\parallel}+\frac{\nu}{v_l}\right)
\end{split}
\end{equation}
where we made the variable change $t = v_l \tau$. We also define the sign of $f_{\perp}$: it is positive when the cross product $\mathbf{v}_l \times \mathbf{f}_{\perp}$ points towards the reader, and negative otherwise. Hence, replacing $f_{\parallel}$ with $-\frac{\nu}{v_l}$ and considering both signs for $f_{\perp}$, we find two frequency vectors that satisfy the Dirac condition, $\mathbf{f_1}$ and $\mathbf{f_2}$, with a norm $f = f_1 = f_2 = \sqrt{f_{\perp}^2+\left(\frac{\nu}{v_l}\right)^2}$ and respective angles $\theta_1 = \theta_0+\theta_{v_l}$ and $\theta_2 = -\theta_0+\theta_{v_l}$, with $\theta_0 = \text{acos}\left(\frac{f_{\parallel}}{f} \right) = \text{acos}\left(-\frac{\nu}{f v_l} \right)$ and $\theta_{v_l}$ the angle between $\mathbf{v}_l$ and the X axis. \eqref{eqphasepsd2} then becomes:
\begin{equation}
\label{eqphasepsd3}
\begin{split}
S_{\phi_1,\phi_2,l} (\nu) = & \frac{1}{v_l} \int_0^{\infty} df_{\perp}\ \Bigg[ \frac{J_1 \left(2\pi R_1 (1-A_{1l}) (K-1) f \right)}{\pi R_1 (1-A_{1l}) (K-1) f} \\
& - \frac{J_1 \left(2\pi R_1 (1-A_{1l}) f \right)}{\pi R_1 (1-A_{1l}) f} \frac{J_1 \left(2\pi R_2 (1-A_{2l}) f \right)}{\pi R_2 (1-A_{2l}) f} \Bigg] \\
& \times \Bigg\{  W_{\phi_l}\left(-\frac{\nu}{v_l},f_{\perp}\right) \exp[-2i\pi \mathbf{f_1} \cdot  \mathbf{s}_l] \\
& +  W_{\phi_l}\left(-\frac{\nu}{v_l},-f_{\perp}\right) \exp[-2i\pi \mathbf{f_2} \cdot  \mathbf{s}_l] \Bigg\}
\end{split}
\end{equation}
By definition, $W_{\phi_l}$ is a radial quantity, so $W_{\phi_l}\left(-\frac{\nu}{v_l},f_{\perp}\right) = W_{\phi_l}\left(-\frac{\nu}{v_l},-f_{\perp}\right) = W_{\phi_l}(f)$. We can thus write the final expression of the piston-removed phase CPSD:
\begin{equation}
\label{eqphasepsdf}
\begin{split}
S_{\phi_1,\phi_2,l} (\nu) = & \frac{1}{v_l} \int_0^{\infty} df_{\perp}\ W_{\phi_l}(f) \Bigg[ \frac{J_1 \left(2\pi R_1 (1-A_{1l}) (K-1) f \right)}{\pi R_1 (1-A_{1l}) (K-1) f} \\
& - \frac{J_1 \left(2\pi R_1 (1-A_{1l}) f \right)}{\pi R_1 (1-A_{1l}) f} \frac{J_1 \left(2\pi R_2 (1-A_{2l}) f \right)}{\pi R_2 (1-A_{2l}) f} \Bigg] \\
& \times \Big\{  \exp[-2i\pi  f s_l\cos(\theta_1 -\theta_{s_l})] \\
& + \exp[-2i\pi  f s_l\cos(\theta_2 -\theta_{s_l})]  \Big\}
\end{split}
\end{equation}
with $\theta_{s_l}$ the angle between $\mathbf{s}_l$ and the X axis. We can also write the CPSD value at the corresponding negative frequency:
\begin{equation}
\label{eqphasepsdneg}
\begin{split}
S_{\phi_1,\phi_2,l} (-\nu) = & \frac{1}{v_l} \int_0^{\infty} df_{\perp}\ W_{\phi_l}(f) \Bigg[ \frac{J_1 \left(2\pi R_1 (1-A_{1l}) (K-1) f \right)}{\pi R_1 (1-A_{1l}) (K-1) f} \\
& - \frac{J_1 \left(2\pi R_1 (1-A_{1l}) f \right)}{\pi R_1 (1-A_{1l}) f} \frac{J_1 \left(2\pi R_2 (1-A_{2l}) f \right)}{\pi R_2 (1-A_{2l}) f} \Bigg] \\
& \times \Big\{  \exp[-2i\pi  f s_l\cos(\theta_3 -\theta_{s_l})] \\
& + \exp[-2i\pi  f s_l\cos(\theta_4 -\theta_{s_l})]  \Big\}
\end{split}
\end{equation}
with $\theta_3 = \pi+\theta_1 = \pi+\theta_0+\theta_{v_l}$ and $\theta_4 = \pi+\theta_2 = \pi-\theta_0+\theta_{v_l}$.
One can easily show that $S_{\phi_1,\phi_2,l} (-\nu)$ is the conjugate of $S_{\phi_1,\phi_2,l} (\nu)$. We then have $\int_{-\infty}^{\infty} d\nu S_{\phi_1,\phi_2,l}(\nu) = \int_0^{\infty} d\nu S'_{\phi_1,\phi_2,l} (\nu) = C_{\phi_1,\phi_2,l}$, with $S'_{\phi_1,\phi_2,l} (\nu) = 2 \Re\left[S_{\phi_1,\phi_2,l} (\nu)\right]$, $\Re[\ ]$ being the real part, that is:
\begin{equation}
\label{eqphasepsdf2}
\begin{split}
S'_{\phi_1,\phi_2,l} (\nu) = & \frac{2}{v_l} \int_0^{\infty} df_{\perp}\ W_{\phi_l}(f) \Bigg[ \frac{J_1 \left(2\pi R_1 (1-A_{1l}) (K-1) f \right)}{\pi R_1 (1-A_{1l}) (K-1) f} \\
& - \frac{J_1 \left(2\pi R_1 (1-A_{1l}) f \right)}{\pi R_1 (1-A_{1l}) f} \frac{J_1 \left(2\pi R_2 (1-A_{2l}) f \right)}{\pi R_2 (1-A_{2l}) f} \Bigg] \\
& \times \Big\{  \cos[2\pi f s_l\cos(\theta_1 -\theta_{s_l})] \\
& + \cos[2\pi f s_l\cos(\theta_2 -\theta_{s_l})]  \Big\}
\end{split}
\end{equation}
Using the classical trigonometry formulas for the combination of sinusoids, we can write \eqref{eqphasepsdf2} as:
\begin{equation}
\begin{split}
S'_{\phi_1,\phi_2,l} (\nu) = & \frac{4}{v_l} \int_0^{\infty} df_{\perp}\ W_{\phi_l}(f) \Bigg[ \frac{J_1 \left(2\pi R_1 (1-A_{1l}) (K-1) f \right)}{\pi R_1 (1-A_{1l}) (K-1) f} \\
& - \frac{J_1 \left(2\pi R_1 (1-A_{1l}) f \right)}{\pi R_1 (1-A_{1l}) f} \frac{J_1 \left(2\pi R_2 (1-A_{2l}) f \right)}{\pi R_2 (1-A_{2l}) f} \Bigg] \\
& \times   \cos[2\pi f s_l\cos(\theta_{v_l} -\theta_{s_l})\cos(\theta_0)] \\
& \times \cos[2\pi f s_l\sin(\theta_{v_l} -\theta_{s_l})\sin(\theta_0)] 
\end{split}
\end{equation}
or equivalently:
\begin{equation}
\label{eqphasepsdf3}
\begin{split}
S'_{\phi_1,\phi_2,l} (\nu) = & \frac{4}{v_l} \cos\left[2\pi\nu \frac{s_l}{v_l}  \cos(\Delta\theta)\right] \int_0^{\infty} df_{\perp}\ W_{\phi_l}(f) \\
& \times \Bigg[ \frac{J_1 \left(2\pi R_1 (1-A_{1l}) (K-1) f \right)}{\pi R_1 (1-A_{1l}) (K-1) f} \\
& - \frac{J_1 \left(2\pi R_1 (1-A_{1l}) f \right)}{\pi R_1 (1-A_{1l}) f} \frac{J_1 \left(2\pi R_2 (1-A_{2l}) f \right)}{\pi R_2 (1-A_{2l}) f} \Bigg] \\
& \times  \cos\left[2\pi f_{\perp} s_l \sin(\Delta\theta)\right] 
\end{split}
\end{equation}
with $\Delta\theta = \theta_{v_l} -\theta_{s_l}$.

\begin{figure}[H]
\centering
\includegraphics[width=\linewidth]{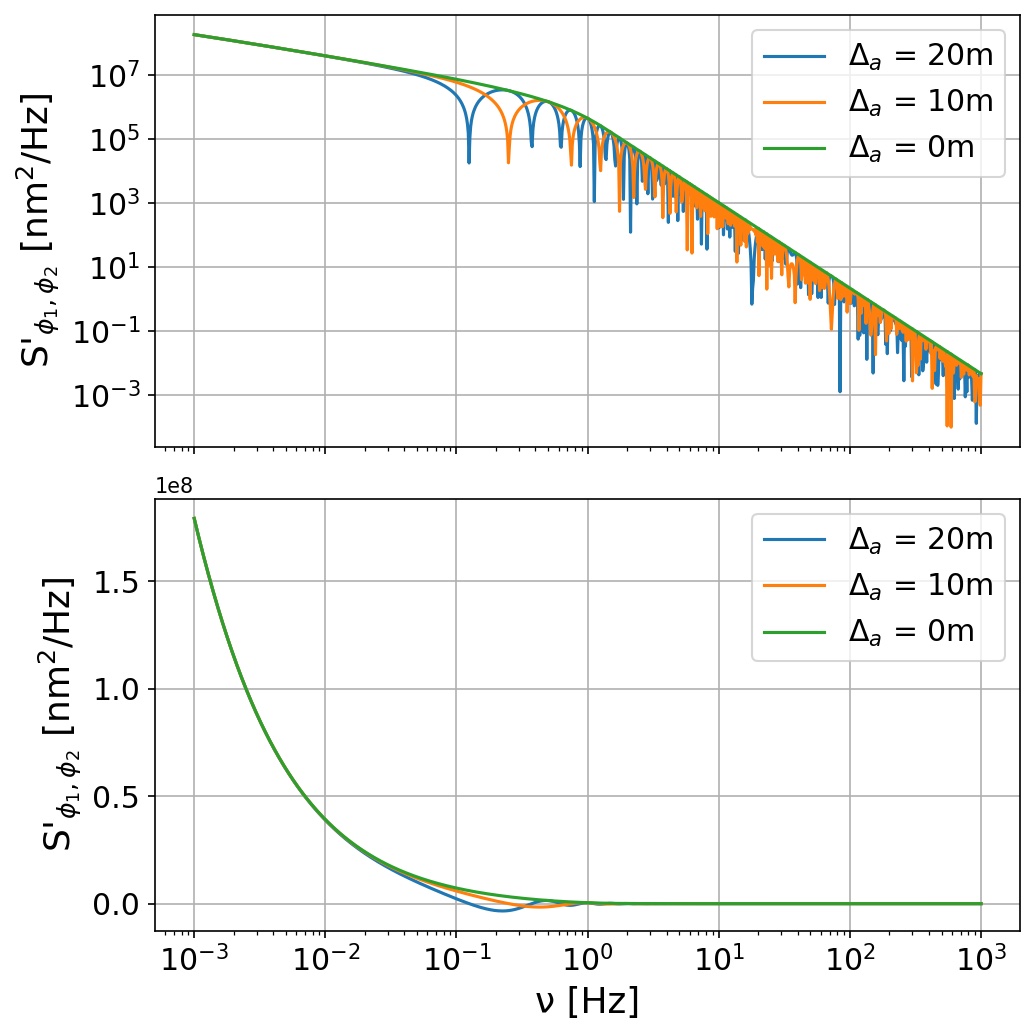}
\caption{CPSD of the piston-removed phase plotted in logarithmic (top) and linear (bottom) scales, for one aperture (green) and two apertures with a separation of 10m (orange) and 20m (blue) along the x-axis. The source is at (0", 0$^{\circ}$, $\infty$) in cylindrical coordinates.}
\label{fig:phase_cpsd}
\end{figure}

In Fig.~\ref{fig:phase_cpsd}, we show the CPSD of the piston-removed phase as derived from \eqref{eqphasepsdf3}. We considered a single-layer turbulent profile with $r_0 = 16cm$, $L_0 = \infty$, $z_l = 10 km$, $v_l = 10 m/s$, $\theta_{v_l} = 0^{\circ}$ and both a single 8m aperture and two 8m apertures looking at one source at infinity. We retrieve the $\nu^{-8/3}$ power law at high frequencies, as shown by Conan \textit{et al.} \cite{conan} for the full turbulent phase (the piston contribution is negligible at high frequencies). At low temporal frequencies, we get a $\nu^{-2/3}$ power law that reflects the major contribution of tip/tilt due to the piston filtering. We also note that the frequencies $\nu_0$ representing the transition from correlation to anti-correlation (and vice versa) show a dependence on the apertures separation. From the formula, we also find a dependence on the wind velocity and $\Delta\theta$. The complete expression is: $\nu_0 = \dfrac{1}{4}\dfrac{v_l}{s_l}\dfrac{1}{cos(\Delta\theta)} (1 + 2k)$, for any integer $k$.

\section{Spatio-temporal cross power spectrum of Zernike coefficients}
\label{secpsd}
We now use the geometry introduced in Fig.~\ref{figgeom2} to develop Whiteley \textit{et al.}'s \cite{whiteley} equation describing the spatio-temporal CPSD of the Zernike coefficients:
\begin{equation}
\begin{split}
S_{a_{1j},a_{2k},l} (\nu) = & \int d\tau\ \int d\mathbf{f}\ W_{\phi_l}(\mathbf{f}) \exp[-2i\pi \mathbf{f} \cdot  \mathbf{s}'_l (\tau)] \\ 
& \times Q_j\left(R_1 (1-A_{1l}) \mathbf{f} \right) Q_k^*\left(R_2 (1-A_{2l}) \mathbf{f} \right) \exp[2i\pi \nu \tau]
\end{split}
\end{equation}
where $a_{1j}$ and $a_{2k}$ are the coefficients representing, respectively, the phases $\phi_1$ and $\phi_2$ in the apertures 1 and 2:
\begin{equation}
a_{1j} = \int d\pmb{\rho}_1 \phi_1(R_1 \pmb{\rho}_1) Z_j(\pmb{\rho}_1) P(\pmb{\rho}_1)
\end{equation}
\begin{equation}
a_{2k} = \int d\pmb{\rho}_2 \phi_2(R_2 \pmb{\rho}_2) Z_k(\pmb{\rho}_2) P(\pmb{\rho}_2)
\end{equation}
and $Q_j(\mathbf{f})$ is the Fourier transform of $Z_j(\pmb{\rho}) P(\pmb{\rho})$:
\begin{equation}
\begin{split}
Q_j(f,\theta) = & i^{m_j} \sqrt{n_j + 1} (-1)^{(n_j-m_j)/2} \sqrt{2}^{1-\delta_{m_j0}} \\
 & \times \frac{J_{n_j + 1}(2\pi f)}{\pi f} \cos \left\{m_j \theta + \frac{\pi}{4} (1-\delta_{m_j0}) [(-1)^j-1] \right\}
\end{split}
\end{equation}
with $n_j$ and $m_j$ the radial and azimuthal orders of $Z_j$ and $\delta_{m_j0}$ the Kronecker delta ($= 1$ if $m_j = 0$, $= 0$ otherwise).\\
Following the reasoning in section \ref{secphasepsd}, we find:
\begin{equation}
\label{eqpsd1}
\begin{split}
S_{a_{1j},a_{2k},l} (\nu) = & \frac{1}{v_l} \int_0^{\infty} df_{\perp}\ W_{\phi_l}(f) \Big[  \exp[-2i\pi \mathbf{f_1} \cdot  \mathbf{s}_l] \\
& \times Q_j\left(R_1 (1-A_{1l}) \mathbf{f_1} \right) Q_k^*\left(R_2 (1-A_{2l}) \mathbf{f_1} \right) \\
& + \exp[-2i\pi \mathbf{f_2} \cdot  \mathbf{s}_l] Q_j\left(R_1 (1-A_{1l}) \mathbf{f_2} \right) \\
& \times Q_k^*\left(R_2 (1-A_{2l}) \mathbf{f_2} \right) \Big]
\end{split}
\end{equation}
We now develop the expressions of $Q_j$ and $Q_k^*$ to find the final CPSD expression:
\begin{equation}
\label{eqpsdf}
\begin{split}
S_{a_{1j},a_{2k},l} (\nu) = & (-1)^{m_k} i^{n_j+n_k} \sqrt{(n_j+1)(n_k+1)} 2^{1-(\delta_{m_j0}+\delta_{m_k0})/2} \\
& \times [v_l \pi^2 R_1 R_2 (1-A_{1l}) (1-A_{2l})]^{-1} \int_0^{\infty} \frac{df_{\perp}}{f^2} \\
& \times W_{\phi_l}(f) J_{n_j+1}\left(2\pi R_1 (1-A_{1l}) f \right) \\
& \times J_{n_k+1}\left(2\pi R_2 (1-A_{2l}) f \right) \\
 & \times \Big[  \exp[-2i\pi f s_l\cos(\theta_1 -\theta_{s_l})] \\
 & \times \cos\left\{m_j \theta_1 + \frac{\pi}{4} (1-\delta_{m_j0}) [(-1)^j-1] \right\} \\
 & \times  \cos\left\{m_k \theta_1 + \frac{\pi}{4} (1-\delta_{m_k0}) [(-1)^k-1] \right\} \\
&  + \exp[-2i\pi f s_l\cos(\theta_2 -\theta_{s_l})] \\
& \times \cos\left\{m_j \theta_2 + \frac{\pi}{4} (1-\delta_{m_j0}) [(-1)^j-1] \right\} \\
 & \times  \cos\left\{m_k \theta_2 + \frac{\pi}{4} (1-\delta_{m_k0}) [(-1)^k-1] \right\} \Big]
\end{split}
\end{equation}
In the specific case of one aperture of radius $R$, one source only at infinity, $j = k$ and a wind along the X axis, we have:
\begin{equation}
\begin{split}
S_{a_j,a_j,l} (\nu) = & \left(n_j+1\right)  \frac{2^{2-\delta_{m_j0}}}{v_l \pi^2 R^2} \int_0^{\infty} \frac{df_{\perp}}{f^2}\ W_{\phi_l}(f) J_{n_j+1}\left(2\pi R f \right)^2 \\
& \times   \cos^2\left\{m_j \theta_0 + \frac{\pi}{4} (1-\delta_{m_j0}) [(-1)^j-1] \right\}
\end{split}
\end{equation}
which is equivalent to the combination of Eqs. (8) and (27) in Conan \textit{et al.} \cite{conan}, with $f_{\perp} = f_y$. 
Keeping the same assumptions, we also derive the PSD of the differential piston between two apertures of same radius. Assuming a homogeneous and isotropic turbulence, one can demonstrate that this PSD is equal to (with $j = k = 1$ for the piston):
\begin{equation}
\begin{split}
    S_{dpist,l}(\nu) = & S_{a_{11},a_{11},l}(\nu)-S_{a_{11},a_{21},l}(\nu)-S_{a_{21},a_{11},l}(\nu) + S_{a_{21},a_{21},l}(\nu) \\
    = & 2 \left\{ S_{a_{11},a_{11},l}(\nu)-\Re[S_{a_{11},a_{21},l}(\nu)]\right\}
\end{split}
\end{equation}
That is:
\begin{equation}
\begin{split}
S_{dpist,l}(\nu) = & \frac{4}{v_l \pi^2 R^2} \int_0^{\infty} \frac{df_{\perp}}{f^2}\ W_{\phi_l}(f) J_{1}\left(2\pi R f \right)^2 \\
& \times \left[  1-\cos\right(2 \pi f s_l\left) \right]
\end{split}
\end{equation}
Hence:
\begin{equation}
    S_{dpist,l}(\nu) = \frac{8}{v_l \pi^2 R^2} \int_0^{\infty} \frac{df_{\perp}}{f^2}\ W_{\phi_l}(f) J_{1}\left(2\pi R f \right)^2 \text{sin}\left(\pi \nu \frac{s_l}{v_l} \right)^2
\end{equation}
which is equivalent to Eq.~(19) in Conan \textit{et al.} \cite{conan} with $s_l = B$.

\begin{figure}[H]
\centering
\includegraphics[width=\linewidth]{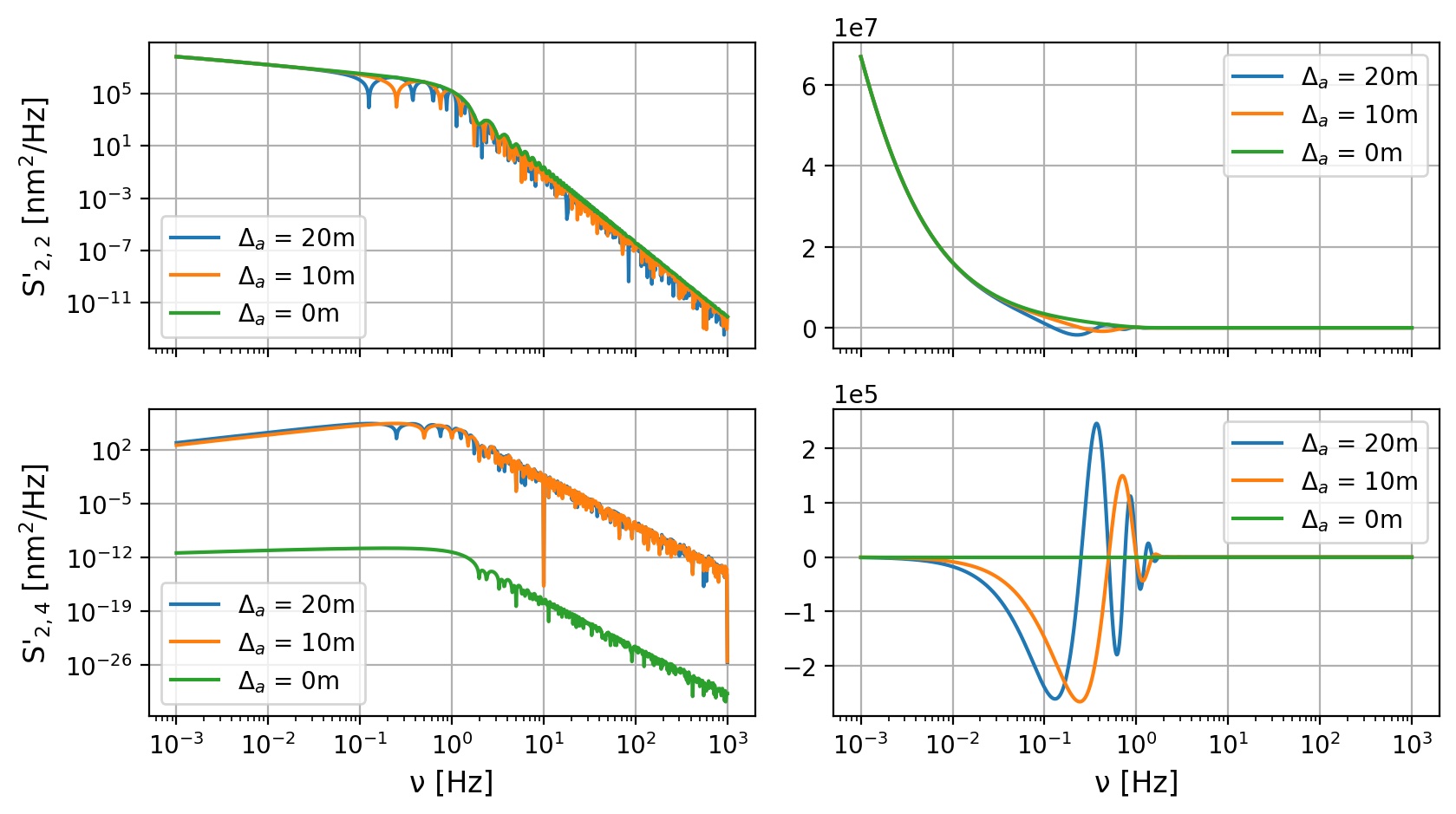}
\caption{CPSD of tip-tip (top) and tip-defocus (bottom), plotted in logarithmic (left) and linear (right) scales. The aperture-source configuration and the turbulence profile are the same as Fig.~\ref{fig:phase_cpsd}.}
\label{fig:tiptip_tipdefocus_cpsd}
\end{figure}

\begin{figure}[H]
\centering
\includegraphics[width=\linewidth]{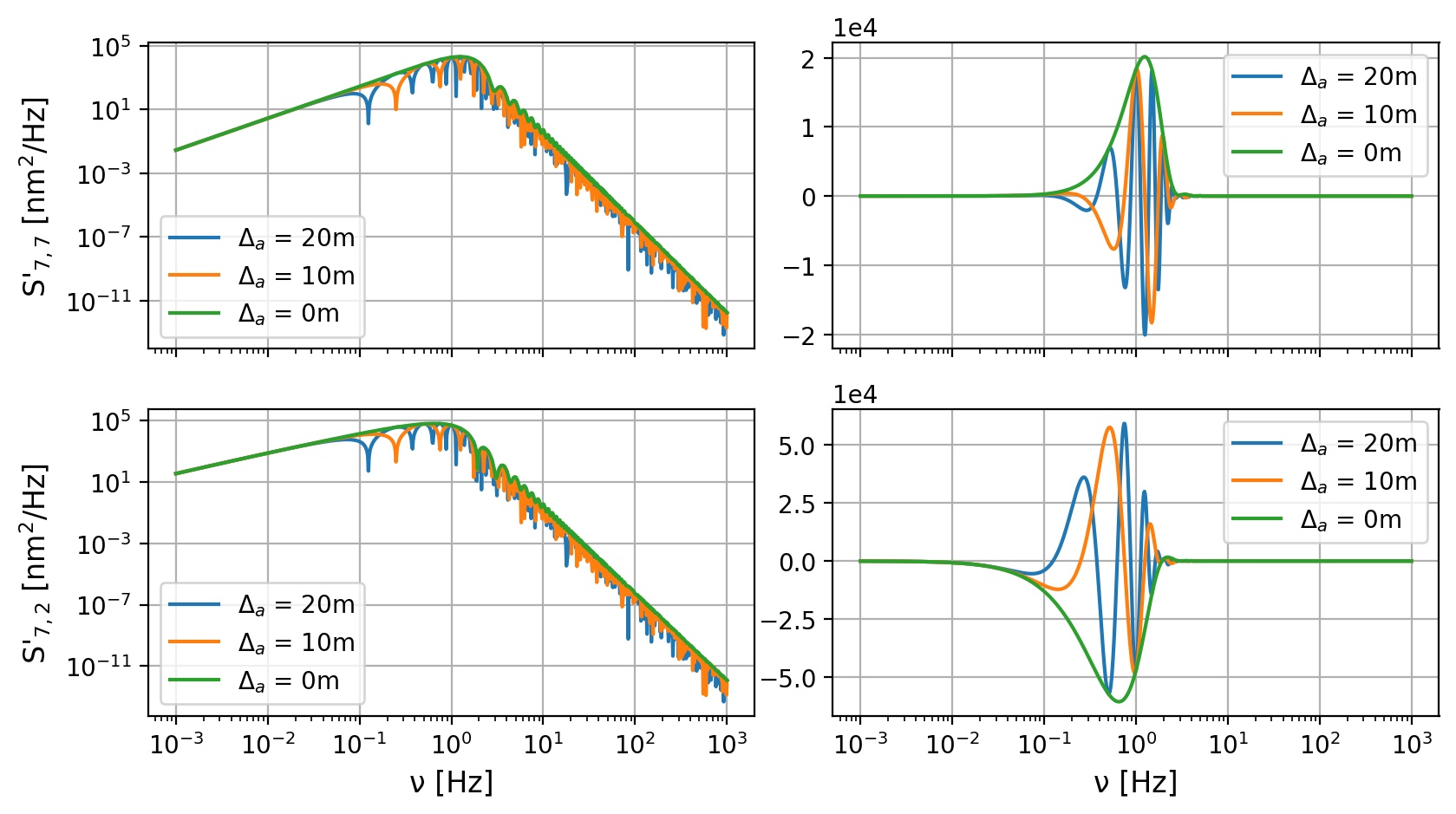}
\caption{CPSD of coma-coma (top) and coma-tip (bottom), plotted in logarithmic (left) and linear (right) scales. The aperture-source configuration and the turbulence profile are the same as Fig.~\ref{fig:phase_cpsd}.}
\label{fig:comacoma_comatip_cpsd}
\end{figure}

As in section \ref{secphasepsd}, we also compute $S'_{a_{1j},a_{2k},l}$, defined in the same way as $S'_{\phi_1,\phi_2,l}$. Indeed, in this case as well, one can verify that $S_{a_{1j},a_{2k},l} (-\nu) = S_{a_{1j},a_{2k},l} (\nu)^*$. The expression of $S'_{a_{1j},a_{2k},l}$ depends on the parity of $n_j+n_k$:
\begin{itemize}
    \item if $n_j+n_k$ is even:
    \begin{equation}
\begin{split}
S'_{a_{1j},a_{2k},l} (\nu) = & (-1)^{m_k} i^{n_j+n_k} \sqrt{(n_j+1)(n_k+1)} \\
& \times 2^{2-(\delta_{m_j0}+\delta_{m_k0})/2} \\
& \times [v_l \pi^2 R_1 R_2 (1-A_{1l}) (1-A_{2l})]^{-1} \\
& \times \int_0^{\infty} \frac{df_{\perp}}{f^2}\ W_{\phi_l}(f) J_{n_j+1}\left(2\pi R_1 (1-A_{1l}) f \right) \\
 & \times J_{n_k+1}\left(2\pi R_2 (1-A_{2l}) f \right) \\
 & \times \Big[  \cos[2\pi f s_l\cos(\theta_1 -\theta_{s_l})] \\
 & \times \cos\left\{m_j \theta_1 + \frac{\pi}{4} (1-\delta_{m_j0}) [(-1)^j-1] \right\} \\
 & \times \cos\left\{m_k \theta_1 + \frac{\pi}{4} (1-\delta_{m_k0}) [(-1)^k-1] \right\} \\
&  + \cos[2\pi f s_l\cos(\theta_2 -\theta_{s_l})] \\
& \times \cos\left\{m_j \theta_2 + \frac{\pi}{4} (1-\delta_{m_j0}) [(-1)^j-1] \right\} \\
 & \times \cos\left\{m_k \theta_2 + \frac{\pi}{4} (1-\delta_{m_k0}) [(-1)^k-1] \right\} \Big]
\end{split}
\end{equation}
\item if $n_j+n_k$ is odd:
    \begin{equation}
\begin{split}
S'_{a_{1j},a_{2k},l} (\nu) = & (-1)^{m_k} i^{n_j+n_k-1} \sqrt{(n_j+1)(n_k+1)} \\
& \times 2^{2-(\delta_{m_j0}+\delta_{m_k0})/2}  \\
 & \times [v_l \pi^2 R_1 R_2 (1-A_{1l}) (1-A_{2l})]^{-1} \\
 & \times \int_0^{\infty} \frac{df_{\perp}}{f^2}\ W_{\phi_l}(f) J_{n_j+1}\left(2\pi R_1 (1-A_{1l}) f \right) \\
 & \times J_{n_k+1}\left(2\pi R_2 (1-A_{2l}) f \right) \\
 & \times \Big[  \sin[2\pi f s_l\cos(\theta_1 -\theta_{s_l})] \\
 & \times \cos\left\{m_j \theta_1 + \frac{\pi}{4} (1-\delta_{m_j0}) [(-1)^j-1] \right\} \\
 & \times \cos\left\{m_k \theta_1 + \frac{\pi}{4} (1-\delta_{m_k0}) [(-1)^k-1] \right\} \\
&  + \sin[2\pi f s_l\cos(\theta_2 -\theta_{s_l})] \\
& \times \cos\left\{m_j \theta_2 + \frac{\pi}{4} (1-\delta_{m_j0}) [(-1)^j-1] \right\} \\
 & \times \cos\left\{m_k \theta_2 + \frac{\pi}{4} (1-\delta_{m_k0}) [(-1)^k-1] \right\} \Big]
\end{split}
\end{equation}
\end{itemize}
We derive the general formula:
\begin{equation}
\label{eqpsd_zern_real}
\begin{split}
S'_{a_{1j},a_{2k},l} (\nu) = & (-1)^{m_k} i^{n_j+n_k+\left[(-1)^{n_j+n_k}-1\right]/2} \sqrt{(n_j+1)(n_k+1)} \\
& \times 2^{2-(\delta_{m_j0}+\delta_{m_k0})/2}  \\
 & \times [v_l \pi^2 R_1 R_2 (1-A_{1l}) (1-A_{2l})]^{-1} \\
 & \times \int_0^{\infty} \frac{df_{\perp}}{f^2}\ W_{\phi_l}(f) J_{n_j+1}\left(2\pi R_1 (1-A_{1l}) f \right) \\
 & \times J_{n_k+1}\left(2\pi R_2 (1-A_{2l}) f \right)  \\
 & \times \Big[  \cos\left\{2\pi f s_l\cos(\theta_1 -\theta_{s_l}) + \frac{\pi}{4} [(-1)^{n_j+n_k}-1] \right\}  \\
 & \times \cos\left\{m_j \theta_1 + \frac{\pi}{4} (1-\delta_{m_j0}) [(-1)^j-1] \right\} \\
 & \times \cos\left\{m_k \theta_1 + \frac{\pi}{4} (1-\delta_{m_k0}) [(-1)^k-1] \right\} \\
&  + \cos\left\{2 \pi f s_l\cos(\theta_2 -\theta_{s_l}) +\frac{\pi}{4} [(-1)^{n_j+n_k}-1] \right\} \\
 & \times \cos\left\{m_j \theta_2 + \frac{\pi}{4} (1-\delta_{m_j0}) [(-1)^j-1] \right\} \\
 & \times \cos\left\{m_k \theta_2 + \frac{\pi}{4} (1-\delta_{m_k0}) [(-1)^k-1] \right\} \Big]
\end{split}
\end{equation}
In Fig.~\ref{fig:tiptip_tipdefocus_cpsd}, we show the CPSDs of tip-tip ($j=2$, $k=2$) and tip-defocus ($j=2$, $k=4$). In Fig.~\ref{fig:comacoma_comatip_cpsd}, we show the CPSDs of coma-coma ($j=8$, $k=8$) and coma-tip ($j=8$, $k=2$). We used the same parameters as in Fig.~\ref{fig:phase_cpsd}. The curves behavior is in agreement with Conan \textit{et al.} \cite{conan}, indeed we retrieve the following power-laws: $\nu^{-17/3}$ at the high frequencies for both $S'_{2,2}$ and $S'_{8,8}$ and, at the low frequencies, $\nu^{-2/3}$ for $S'_{2,2}$ and $\nu^{+2}$ for $S'_{8,8}$. The CPSDs zeros are at $\nu_0 = \dfrac{1}{2} \dfrac{v_l}{s_l} \Big\{\dfrac{1}{2} - \dfrac{1}{4} \big[(-1)^{n_j + n_k} - 1\big] + k\Big\}$, for any integer $k$. This expression is valid only for $\Delta\theta = 0$, as we were unable to derive a general expression for $\Delta\theta \neq 0$.

\section{Numerical applications}
\label{numver}
In this section, we propose an analytical method that requires the CPSDs to estimate the wavefront residuals that are left by a SCAO system sensing the turbulence-induced distortions from an off-axis reference star. We focus on residuals that are only due to anisoplanatism and temporal filtering of the AO control. We neglect other sources of error (fitting error, aliasing error, wavefront sensor noise error, ...), as they are beyond the scope of our application.\\
Though some approaches consider the correlation between anisoplanatism and temporal errors \cite{ellerbroek2005linear,jolissaint,clare2006adaptive,correia2017modeling}, these two terms are often studied separately \cite{madec,clenet,gendron2014novel,sandler,neichel2009tomographic,rigaut1998analytical}.
The former is usually evaluated from the covariances of the turbulent phase \cite{sandler,clenet,chassat}. The latter is determined from the AO control filtering of the temporal Power Spectral Density (PSD) of the turbulence (Eq. (6.36) in Madec \cite{madec}) and it represents the error left on the guide star. Through the CPSDs though, we are able to study the temporal and the spatial errors together and estimate anisoplanatism as affected by the temporal filtering of the adaptive optics correction. In the following, we develop these computations and we show the difference between the anisoplanatism error as computed through the covariances or using the CPSDs. We consider either a single-aperture or a two-aperture interferometric telescope.

\subsection{Time-filtered anisoplanatism for a single-aperture telescope}
\begin{figure}[H]
\centering
\includegraphics[width=\linewidth]{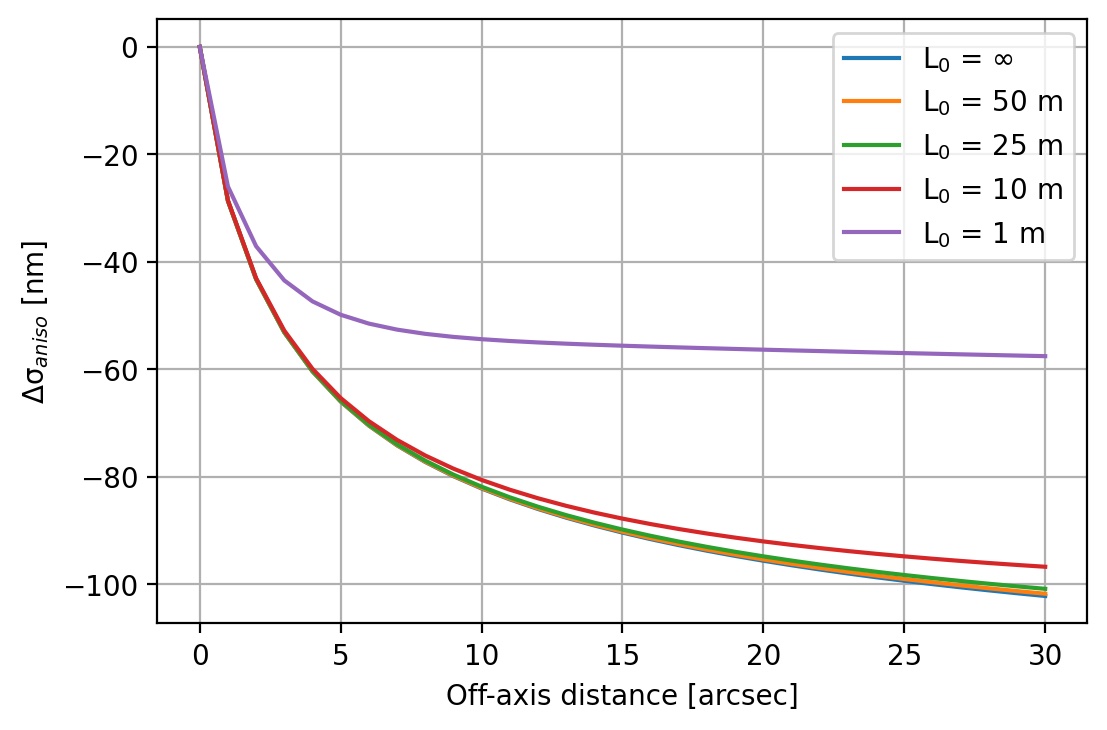}
\caption{Difference between the anisoplanatism error computed through the CPSD or with the covariance method, as a function of the angular off-axis distance of the NGS and for several values of the turbulence outer scale. The assumed telescope is the LBT.}
\label{fig:one_aperture_phase}
\end{figure}

We consider an aperture observing a target on axis and sensing the phase aberrations from an off-axis Natural Guide Star (NGS).\\
The residual phase on target is given by the difference between the turbulent phase on target and the correction phase estimated from the NGS, which we write as:
\begin{equation}
    \varphi_c (\nu) = \varphi_n (\nu) - RTF(\nu) \varphi_n (\nu)
\end{equation}
where $\varphi_n$ is the turbulent phase in the direction of the NGS and where we assumed that no noise is introduced in the AO loop. $RTF$ is the Rejection Transfer Function that, together with the Noise Transfer Function, characterizes the AO control. In the following, we assume a simple integrator controller with $RTF(z) = \frac{1 - z^{-1}}{1 - z^{-1} + g z^{-d}}$, $NTF(z) = - \frac{g z^{-d}}{1 - z^{-1} + g z^{-d}}$, where $g$ is the scalar gain, $d$ is the total delay in frames and $z$ is the temporal-frequency vector defined as $z = e^{2 i \pi \nu / \nu_{loop}}$, with $\nu_{loop}$ the AO loop's frequency \cite{conan1a2011integral}. Note that we consider a control law that is applied to the whole piston-filtered phase, which is an approximation. In reality, the deformable mirror cannot correct all the high spatial frequency content of the turbulent phase and the correction strategy might foresee different gains for different wavefront modes (e. g. a higher gain on tip/tilt to deal with vibrations). \\
The residual phase on target is then:
\begin{equation}
    \label{eq:res_phase_target}
    \begin{split}
        \varphi_{res,target} (\nu) &= \varphi_t (\nu) -  \varphi_c (\nu) \\
        &= \varphi_t (\nu) - \varphi_n (\nu) + RTF(\nu) \varphi_n (\nu) \\
        &= RTF(\nu) \varphi_t (\nu) + NTF(\nu) \big( \varphi_n (\nu) - \varphi_t (\nu) \big)
    \end{split}
\end{equation}
where $\varphi_t$ is the turbulent phase in the direction of the target and where we used the following relationship for the transfer functions: $RTF(\nu) - NTF(\nu) = 1$. It is worth noting that anisoplanatism, described in the equation by the difference between $\varphi_n$ and $\varphi_t$, is filtered as a noise by the AO loop.\\
From \eqref{eq:res_phase_target}, we can compute the temporal PSD of the phase residuals on target as:
\begin{equation}
    \label{eq:scao_res_cpsd}
    \begin{split}
        S_{res,target}(\nu) &= \big\langle \varphi_{res,target}(\nu) \ \varphi_{res,target}(\nu)^{\dagger} \big\rangle \\
        &=|RTF(\nu)|^2 S_{turb}(\nu) + 2 Re \Big[NTF(\nu) \big(S_{n,t}(\nu) \\
        &- S_{turb}(\nu) \big)\Big]
    \end{split}
\end{equation}
where $\big\langle \big\rangle$ is the mean, $S_{turb}$ is the PSD of the turbulence and $S_{n,t}$ is the CPSD between the phase on the NGS and the phase on target. By assuming homogeneous and isotropic atmospheric turbulence, we considered $S_{t,t}=S_{n,n}=S_{turb}$ and $S_{t,n}=S_{n,t}$.\\
The first term of the equation represents the residual PSD left in the direction of the NGS \cite{madec}. The last two terms represent the residual PSD due to anisoplanatism as filtered by the AO control. If integrated with respect to the temporal frequencies, it provides the anisoplanatism error that is generally computed through the spatial covariances of the phase as \cite{chassat,sandler,clenet}:
\begin{equation}
    \sigma^2_{aniso} = 2 (\sigma^2_{t,t} - \sigma^2_{n,t})
\end{equation}
In Fig.~\ref{fig:one_aperture_phase}, we investigate the difference between the anisoplanatism error computed through the CPSDs or through the covariances. We show the results for the phase and we use \eqref{eqphasepsdf3} to compute the CPSDs, in order to limit the integration to the temporal frequency range [0,$\infty$) and thus gain computation time. As a case study, we considered the LBT \cite{hill} observing with one of the two 8.2m pupils and compensating the turbulence-induced distortions through a SCAO correction characterized by $\nu_{loop}$ = 500Hz, $g$ = 0.2 and $d$ = 2. The turbulence profile we used to compute the CPSDs is a four-layer profile taken from Agapito \textit{et al.} \cite{agapito2014adaptive}. The parameters are shown in Table~\ref{table:cn2_profile}. We computed every CPSD as sum of the single-layer CPSDs, assuming that the phase perturbations at each layer are not correlated.\\
As expected, the anisoplanatism error computed through the CPSD method is smaller than the one computed through the covariances, as it includes the temporal filtering by the AO control. We note that this behavior is valid within the whole range of $L_0$, which is between limit values of 1m and $\infty$. For the typical values (10m - 50m), measuring the anisoplanatism error through the covariances leads to an overestimation of $\sim$ 40nm at $\sim$ 2" off axis. It is also worth noting that the filtered anisoplanatism error given by the presented approach can be used to compute the isoplanatic patch after AO correction, defined for example in Agapito \textit{et al.}\cite{agapito2014adaptive} as $\theta_N$, in a much faster way than with end-to-end simulations.

\begin{table}[H]
\small
\centering
\begin{tabular}{ |c|c|c|c|c| } 
 \hline
 \textit{Height [m]} & 103 & 725 & 2637 & 11068 \\
 \hline
 \textit{$C_N^2$ fraction} & 0.70 & 0.06 & 0.14 & 0.10 \\
 \hline
 \textit{Wind speed [m/s]} & 2 & 4 & 6 & 25 \\ 
 \hline
 \textit{Seeing [arcsec]} & \multicolumn{4}{c|}{0.66} \\
 \hline
 \textit{Zenith angle [$^\circ$]} & \multicolumn{4}{c|}{40} \\
 \hline
\end{tabular}
\caption{Parameters of the atmospheric turbulence profile. The seeing and layer altitudes are given at zenith and are scaled with respect to the airmass in the simulation.}
\label{table:cn2_profile}
\end{table}

\subsection{Time-filtered anisoplanatism for an interferometric telescope}
\begin{figure}[H]
\centering
\includegraphics[width=\linewidth]{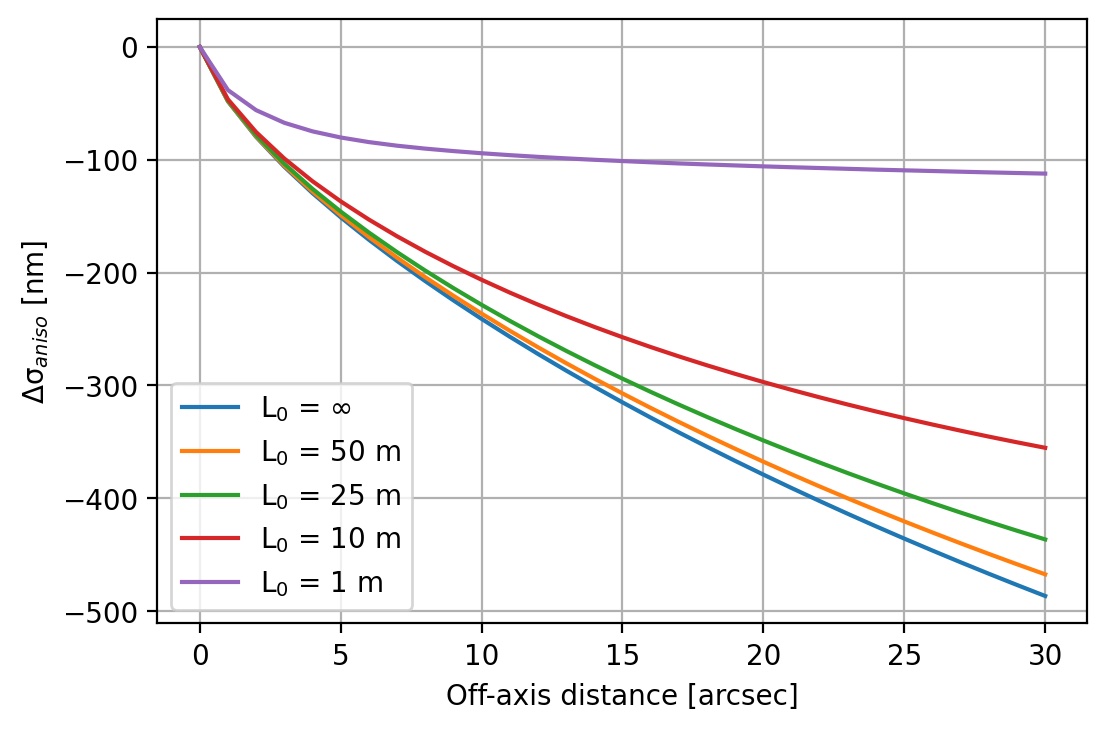}
\caption{Difference between the anisoplanatism error computed through the CPSD or with the covariance method, as a function of the angular off-axis distance of the NGS and for several values of the turbulence outer scale. The assumed interferometric telescope is the LBTI.}
\label{fig:two_aperture_phase}
\end{figure}

We now consider a two-aperture interferometric telescope. The off-axis NGS is needed to sense the differential phase between the two sides of the telescope, that is the signal to be minimized in interferometric observations.\\
In this case, we have to determine the residual PSD from the difference between the residual phases on the two sides of the interferometer. Thus, we define the temporal PSD in the direction of the target as:
\begin{equation}
    \label{eq:interf_res_cpsd}
    \begin{split}
        S_{res,target} (\nu) &= \Big\langle \big(\varphi_{res_{t_1}}(\nu) - \varphi_{res_{t_2}}(\nu)\big) \ \big(\varphi_{res_{t_1}}(\nu) - \varphi_{res_{t_2}}(\nu)\big)^{\dagger}\Big\rangle \\
        &= 2 \bigg\{ |RTF(\nu)|^2 \big( S_{turb}(\nu) - S_{n_1,n_2}(\nu) \big) \\
        & + 2 Re \Big[ NTF(\nu) \big( S_{n_1,t_1}(\nu) - S_{turb}(\nu) \big) \Big]\\
        &+ Re \Big[ NTF(\nu) \big( 2 S_{n_1,n_2}(\nu) - S_{n_1,t_2}(\nu) - S_{n_2,t_1}(\nu) \big) \Big] \bigg\}
    \end{split}
\end{equation}
where $\varphi_{res_{t_1}}$ and $\varphi_{res_{t_2}}$ are the residual phases on the first and the second aperture of the interferometer, both given by \eqref{eq:res_phase_target}, $S_{turb}$ is the PSD of the turbulent phase in a given line of sight seen by a single aperture and $S_{x_i,y_j}$ is the CPSD between the turbulent phases seen from two directions $(x,y)$ by two apertures $(i,j)$. Here, $x$ and $y$ can either be $n$ for the NGS or $t$ for the target.
By assuming homogeneous and isotropic atmospheric turbulence, we used the following relationships for the CPSDs: $S_{t_1,t_1}=S_{t_2,t_2}=S_{n_1,n_1}=S_{n_2,n_2}=S_{turb}$, $S_{t_1,n_1}=S_{t_2,n_2}$, $S_{t_1,t_2}=S_{t_2,t_1}=S_{n_1,n_2}=S_{n_2,n_1}$, $S_{t_1,n_2}=S_{n_2,t_1}$. 
The two AO systems equipping each side of the interferometer work independently from one another, but as they see the same atmospheric conditions and the same star, we assumed that they have the same control law (e. g. the same integrator gain and delay), hence they are characterized by the same \textit{RTF} and \textit{NTF}.
As in the single-aperture case, we note that the first two terms represent the residual PSD left in the direction of the NGS, while the last five terms, if integrated on the temporal frequencies, represent the anisoplanatism error as filtered by the AO loops. This last source of error is usually computed through the spatial covariances of the phase, as we find in Esposito \textit{et al.} \cite{esposito}. The formula, presented for the differential piston errors but still valid for the differential phase in general, is:
\begin{equation}
    \sigma^2_{aniso}= 2 (2 \sigma^2_{turb} - 2 \sigma^2_{t_1,n_1} - 2 \sigma^2_{n_1,n_2} + \sigma^2_{t_1,n_2} + \sigma^2_{t_2,n_1})
\end{equation}
In Fig.~\ref{fig:two_aperture_phase}, we show the difference between the anisoplanatism error computed through the CPSDs or through the covariances. As an interferometric telescope, we considered the LBTI, that is characterized by two 8.2m pupils and a center-to-center distance of 14.4m. We used the same parameters as in the previous paragraph for the AO loop and the turbulence profile. As in the single-aperture case, we note that the AO temporal filtering has a significant effect in reducing the contribution of anisoplanatism, with a difference greater than 200 nm at 10" off axis for typical $L_0$ values.

This application shows that the two beams of a small-baseline interferometer are likely to be highly correlated. This correlation between adjacent beams can be of interest for the study of segmented telescopes. For example, the Giant  Magellan Telescope (GMT) \cite{hinz}, that has a primary composed of 7 segments and a deformable secondary segmented in the same way, could be considered as a 7-aperture interferometer, since each pair primary segment - secondary segment is equivalent to one side of the LBT. The question is then: is it better to make each segment work independently from the others as in an interferometer, and then adjust the differential pistons, or to consider the full pupil in the control scheme, using global wavefront modes? We do not intend to answer here, as this topic would deserve a publication on its own, but the presented framework should be of great help for such a study.

\section{Conclusion}
 In this paper, we derived analytical formulas for the 
 temporal 
 Cross Power Spectral Densities of the turbulent phases in a configuration with two apertures looking at two distinct sources. We considered either the piston-removed phase or the phase decomposition on Zernike modes. The general geometry allows to cover a wide range of applications in the field of astronomical observations with ground-based telescopes equipped with adaptive optics. 
 Indeed, the presented framework offers the possibility to directly take into account the following aspects altogether in a single formula, while allowing the application of a time filtering: distinct apertures of different size, distinct sources at finite or infinite distance, direction of the wind. It is also focused on temporal frequencies, in contrast with many spatial-frequency-based methods, hence it provides access to a different dimension of the AO performance evaluation. 
 The spatio-temporal behavior of the turbulence-induced distortions can be exploited as a useful tool to estimate the performance of both single-aperture and interferometric telescopes provided with classical and new-generation adaptive optics systems. In this context, we provided 
 an example 
 of numerical application where we used the CPSDs to estimate the wavefront residuals of a SCAO system due to anisoplanatism, taking into account the effect of the adaptive optics control on this source of error. We found that anisoplanatism is filtered as a noise by the AO loop. As a consequence, not considering the AO temporal filtering of anisoplanatism (that is, evaluating this source of error through the spatial covariances of the phase) brings to an overestimation of this error term.
 
\begin{backmatter}

\bmsection{Funding} Premiale ADONI (Cram 1.05.06.07) chiave MAORY.
\bmsection{Disclosures} The authors declare no conflicts of interest.
\bmsection{Data availability} No data were generated or analyzed in the presented research.

\end{backmatter}


\bibliography{temp_psd_cov_zern}

\begin{thebibliography}{10}
\newcommand{\enquote}[1]{``#1''}

\bibitem{noll}
R.~J. Noll, \enquote{Zernike polynomials and atmospheric turbulence,}
  {\protect\JournalTitle{{J. Opt. Soc. Am.}}} \textbf{66} (1976).

\bibitem{sasiela}
R.~J. Sasiela, \enquote{Wave-front correction by one or more synthetic
  beacons,} {\protect\JournalTitle{{J. Opt. Soc. Am. A}}} \textbf{11}, 379--393
  (1994).

\bibitem{conan}
J.-M. Conan, G.~Rousset, and P.-Y. Madec, \enquote{Wave-front temporal spectra
  in high-resolution imaging through turbulence,} {\protect\JournalTitle{{J.
  Opt. Soc. Am. A}}} \textbf{12}, 1559--1570 (1995).

\bibitem{rconan}
R.~Conan, \enquote{{Mean-square residual error of a wavefront after propagation
  through atmospheric turbulence and after correction with Zernike
  polynomials},} {\protect\JournalTitle{{J. Opt. Soc. Am. A}}} \textbf{25},
  526--536 (2008).

\bibitem{navarro}
R.~Navarro, J.~Arines, and R.~Rivera, \enquote{{Direct and inverse discrete
  Zernike transform},} {\protect\JournalTitle{Opt. Express}} \textbf{17},
  24269--24281 (2009).

\bibitem{roddier}
F.~Roddier, M.~Northcott, J.~Graves, D.~McKenna, and D.~Roddier,
  \enquote{{One-dimensional spectra of turbulence-induced Zernike aberrations:
  time-delay and isoplanicity error in partial adaptive compensation},}
  {\protect\JournalTitle{{J. Opt. Soc. Am. A}}} \textbf{10}, 957--965 (1993).

\bibitem{nroddier}
N.~A. Roddier, \enquote{{Atmospheric wavefront simulation using Zernike
  polynomials},} {\protect\JournalTitle{Opt. Eng.}} \textbf{29}, 1174--1181
  (1990).

\bibitem{gendron}
E.~Gendron and G.~Rousset, \enquote{{Temporal analysis of aliasing in
  Shack-Hartmann wave-front sensing},} {\protect\JournalTitle{Proc. SPIE}}
  (2012).

\bibitem{negro}
J.~E. Negro, \enquote{Subaperature optical system testing,}
  {\protect\JournalTitle{Appl. Opt.}} \textbf{23}, 1921--1930 (1984).

\bibitem{molodij}
G.~Molodij and G.~Rousset, \enquote{{Angular correlation of Zernike polynomials
  for a laser guide star in adaptive optics},} {\protect\JournalTitle{{J. Opt.
  Soc. Am. A}}} \textbf{14}, 1949--1966 (1997).

\bibitem{hu}
P.~Hu, J.~Stone, and T.~Stanley, \enquote{{Application of Zernike polynomials
  to atmospheric propagation problems},} {\protect\JournalTitle{{J. Opt. Soc.
  Am. A}}} \textbf{6}, 1595--1608 (1989).

\bibitem{takato}
N.~Takato and I.~Yamaguchi, \enquote{{Spatial correlation of Zernike
  phase-expansion coefficients for atmospheric turbulence with finite outer
  scale},} {\protect\JournalTitle{{J. Opt. Soc. Am. A}}} \textbf{12}, 958--963
  (1995).

\bibitem{whiteley}
M.~R. Whiteley, M.~C. Roggemann, and B.~M. Welsh, \enquote{{{Temporal
  properties of the Zernike expansion coefficients of turbulence-induced phase
  aberrations for aperture and source motion}},} {\protect\JournalTitle{{J.
  Opt. Soc. Am. A}}} \textbf{15}, 993--1005 (1998).

\bibitem{born}
M.~Born and E.~Wolf, \emph{Principles of optics: electromagnetic theory of
  propagation, interference and diffraction of light} (Elsevier, 2013).

\bibitem{hogge}
C.~B. Hogge and R.~R. Butts, \enquote{{Frequency spectra for the geometric
  representation of wavefront distortions due to atmospheric turbulence},}
  {\protect\JournalTitle{{IEEE Transactions on Antennas and Propagation}}}
  \textbf{24}, 144--154 (1976).

\bibitem{pinna2016soul}
E.~Pinna, S.~Esposito, P.~Hinz, G.~Agapito, M.~Bonaglia, A.~Puglisi,
  M.~Xompero, A.~Riccardi, R.~Briguglio, C.~Arcidiacono, L.~Carbonaro, L.~Fini,
  M.~Montoya, and O.~Durney, \enquote{{SOUL}: the single conjugated adaptive
  optics upgrade for {LBT},} in \emph{Adaptive Optics Systems V,}  vol. 9909
  (International Society for Optics and Photonics, 2016), p. 99093V.

\bibitem{petit2016saxo}
C.~Petit, J.-F. Sauvage, A.~Costille, T.~Fusco, D.~Mouillet, J.-L. Beuzit,
  K.~Dohlen, M.~E. Kasper, M.~S. Valles, C.~Soenke, A.~Baruffolo, B.~Salasnich,
  S.~Rochat, E.~Fedrigo, P.~Baudoz, E.~Hugot, A.~Sevin, D.~Perret, F.~Wildi,
  M.~Downing, P.~Feautrier, P.~Puget, A.~Vigan, J.~O’Neal, J.~H.~V. Girard,
  D.~Mawet, H.~M. Schmid, and R.~Roelfsema, \enquote{{SAXO}: the extreme
  adaptive optics system of {SPHERE (I)} system overview and global laboratory
  performance,} {\protect\JournalTitle{Journal of Astronomical Telescopes,
  Instruments, and Systems}} \textbf{2}, 025003 (2016).

\bibitem{neichel2010gemini}
B.~{Neichel}, F.~{Rigaut}, M.~{Bec}, M.~{Boccas}, F.~{Daruich},
  C.~{D'Orgeville}, V.~{Fesquet}, R.~{Galvez}, A.~{Garcia-Rissmann},
  G.~{Gausachs}, M.~{Lombini}, G.~{Perez}, G.~{Trancho}, V.~{Upadhya}, and
  T.~{Vucina}, \enquote{The {Gemini MCAO System GeMS}: nearing the end of a
  lab-story,} in \emph{Adaptive Optics Systems II,}  vol. 7736 (International
  Society for Optics and Photonics, 2010), p. 773606.

\bibitem{stuik2006galacsi}
R.~Stuik, R.~Bacon, R.~Conzelmann, B.~Delabre, E.~Fedrigo, N.~Hubin,
  M.~Le~Louarn, and S.~Str{\"o}bele, \enquote{{GALACSI}--the ground layer
  adaptive optics system for {MUSE},} {\protect\JournalTitle{New Astronomy
  Reviews}} \textbf{49}, 618--624 (2006).

\bibitem{diolaiti}
E.~Diolaiti, P.~Ciliegi, R.~Abicca, G.~Agapito, C.~Arcidiacono, A.~Baruffolo,
  M.~Bellazzini, V.~Biliotti, M.~Bonaglia, G.~Bregoli, R.~Briguglio,
  O.~Brissaud, L.~Busoni, L.~Carbonaro, A.~Carlotti, E.~Cascone, J.-J. Correia,
  F.~Cortecchia, G.~Cosentino, V.~D. Caprio, M.~de~Pascale, A.~D. Rosa, C.~D.
  Vecchio, A.~Delboulbé, G.~D. Rico, S.~Esposito, D.~Fantinel, P.~Feautrier,
  C.~Felini, D.~Ferruzzi, L.~Fini, G.~Fiorentino, I.~Foppiani, M.~Ghigo,
  C.~Giordano, E.~Giro, L.~Gluck, F.~Hénault, L.~Jocou, F.~Kerber, P.~L.
  Penna, S.~Lafrasse, M.~Lauria, E.~le~Coarer, M.~L. Louarn, M.~Lombini,
  Y.~Magnard, E.~Maiorano, F.~Mannucci, M.~Mapelli, E.~Marchetti, D.~Maurel,
  L.~Michaud, G.~Morgante, T.~Moulin, S.~Oberti, G.~Pareschi, M.~Patti,
  A.~Puglisi, P.~Rabou, R.~Ragazzoni, S.~Ramsay, A.~Riccardi, S.~Ricciardi,
  M.~Riva, S.~Rochat, F.~Roussel, A.~Roux, B.~Salasnich, P.~Saracco,
  L.~Schreiber, M.~Spavone, E.~Stadler, M.-H. Sztefek, N.~Ventura,
  C.~Vérinaud, M.~Xompero, A.~Fontana, and F.~M. Zerbi, \enquote{{MAORY:
  adaptive optics module for the E-ELT},} {\protect\JournalTitle{Proc. SPIE}}
  (2016).

\bibitem{neichel}
B.~Neichel, T.~Fusco, J.-F. Sauvage, C.~Correia, K.~Dohlen, K.~El-Hadi,
  L.~Blanco, N.~Schwartz, F.~Clarke, N.~A. Thatte, M.~Tecza, J.~Paufique,
  J.~Vernet, M.~L. Louarn, P.~Hammersley, J.-L. Gach, S.~Pascal, P.~Vola,
  C.~Petit, J.-M. Conan, A.~Carlotti, C.~Vérinaud, H.~Schnetler, I.~Bryson,
  T.~Morris, R.~Myers, E.~Hugot, A.~M. Gallie, and D.~M. Henry, \enquote{{The
  adaptive optics modes for HARMONI: from Classical to Laser Assisted
  Tomographic AO},} {\protect\JournalTitle{Proc. SPIE}}  (2016).

\bibitem{herriot}
G.~Herriot, D.~Andersen, J.~Atwood, C.~Boyer, A.~Beauvillier, P.~Byrnes,
  R.~Conan, B.~Ellerbroek, J.~Fitzsimmons, L.~Gilles, P.~Hickson, A.~Hill,
  K.~Jackson, O.~Lardière, J.~Pazder, T.~Pfrommer, V.~Reshetov, S.~Roberts,
  J.-P. Véran, L.~Wang, and I.~Wevers, \enquote{{NFIRAOS: TMT's facility
  adaptive optics system},} {\protect\JournalTitle{Proc. SPIE}}  (2010).

\bibitem{hinz}
P.~M. Hinz, A.~Bouchez, M.~Johns, S.~Shectman, M.~Hart, B.~McLeod, and
  P.~McGregor, \enquote{{The GMT adaptive optics system},}
  {\protect\JournalTitle{Proc. SPIE}}  (2010).

\bibitem{mcdermid}
R.~McDermid, \enquote{{MAVIS: A new MCAO-Assisted Visible Imager and
  Spectrograph for the Very Large Telescope},} in \emph{Linking Galaxies from
  the Epoch of Initial Star Formation to Today,}  (2019).

\bibitem{conan2012giant}
R.~Conan, F.~Bennet, A.~H. Bouchez, M.~A. van Dam, B.~Espeland, W.~Gardhouse,
  C.~d'Orgeville, S.~Parcell, P.~Piatrou, I.~Price, F.~Rigaut, G.~Trancho, and
  K.~Uhlendorf, \enquote{The {Giant Magellan Telescope} laser tomography
  adaptive optics system,} in \emph{Adaptive Optics Systems III,}  vol. 8447
  (International Society for Optics and Photonics, 2012), p. 84473P.

\bibitem{madec}
P.-Y. Madec, \enquote{Control techniques,} in \emph{Adaptive Optics in
  Astronomy,}  (Cambridge U. Press, Cambridge, UK, 1999), pp. 131--154.

\bibitem{clenet}
Y.~Clénet, E.~Gendron, D.~Gratadour, G.~Rousset, and F.~Vidal,
  \enquote{{Anisoplanatism effect on the {E-ELT SCAO} point spread function. A
  preserved coherent core across the field},} {\protect\JournalTitle{Astronomy
  and Astrophysics}} \textbf{583} (2015).

\bibitem{gendron2014novel}
E.~Gendron, A.~Charara, A.~Abdelfattah, D.~Gratadour, D.~Keyes, H.~Ltaief,
  C.~Morel, F.~Vidal, A.~Sevin, and G.~Rousset, \enquote{A novel fast and
  accurate pseudo-analytical simulation approach for {MOAO},} in \emph{Adaptive
  Optics Systems IV,}  vol. 9148 (International Society for Optics and
  Photonics, 2014), p. 91486L.

\bibitem{sandler}
D.~G. Sandler, S.~Stahl, J.~R.~P. Angel, M.~Lloyd-Hart, and D.~McCarthy,
  \enquote{{Adaptive optics for diffraction-limited infrared imaging with 8-m
  telescopes},} {\protect\JournalTitle{Journal of the Optical Society of
  America A}} \textbf{11}, 925--945 (1994).

\bibitem{neichel2009tomographic}
B.~Neichel, T.~Fusco, and J.-M. Conan, \enquote{Tomographic reconstruction for
  wide-field adaptive optics systems: {Fourier} domain analysis and fundamental
  limitations,} {\protect\JournalTitle{JOSA A}} \textbf{26}, 219--235 (2009).

\bibitem{rigaut1998analytical}
F.~J. Rigaut, J.-P. V{\'e}ran, and O.~Lai, \enquote{Analytical model for
  {Shack-Hartmann-based} adaptive optics systems,} in \emph{Adaptive Optical
  System Technologies,}  vol. 3353 (International Society for Optics and
  Photonics, 1998), pp. 1038--1048.

\bibitem{ellerbroek2005linear}
B.~L. Ellerbroek, \enquote{Linear systems modeling of adaptive optics in the
  spatial-frequency domain,} {\protect\JournalTitle{JOSA A}} \textbf{22},
  310--322 (2005).

\bibitem{jolissaint}
L.~Jolissaint, \enquote{Synthetic modeling of astronomical closed loop adaptive
  optics,} {\protect\JournalTitle{Journal of the European Optical Society -
  Rapid publications}} \textbf{5} (2010).

\bibitem{clare2006adaptive}
R.~M. Clare, B.~L. Ellerbroek, G.~Herriot, and J.-P. V{\'e}ran,
  \enquote{Adaptive optics sky coverage modeling for extremely large
  telescopes,} {\protect\JournalTitle{Applied optics}} \textbf{45}, 8964--8978
  (2006).

\bibitem{correia2017modeling}
C.~M. Correia, C.~Z. Bond, J.-F. Sauvage, T.~Fusco, R.~Conan, and P.~L.
  Wizinowich, \enquote{Modeling astronomical adaptive optics performance with
  temporally filtered {Wiener} reconstruction of slope data,}
  {\protect\JournalTitle{JOSA A}} \textbf{34}, 1877--1887 (2017).

\bibitem{hinz_lbti16}
P.~M. Hinz, D.~Defrère, A.~Skemer, V.~Bailey, J.~Stone, E.~Spalding, A.~Vaz,
  E.~Pinna, A.~Puglisi, S.~Esposito, M.~Montoya, E.~Downey, J.~Leisenring,
  O.~Durney, W.~Hoffmann, J.~Hill, R.~Millan-Gabet, B.~Mennesson, W.~Danchi,
  K.~Morzinski, P.~Grenz, M.~Skrutskie, and S.~Ertel, \enquote{{Overview of
  LBTI: a multipurpose facility for high spatial resolution observations},}
  {\protect\JournalTitle{Proc. SPIE}}  (2016).

\bibitem{chassat}
F.~Chassat, G.~Rousset, and J.~Primot, \enquote{{Theoretical and experimental
  evaluation of isoplanatic patch size for adaptive optics},}
  {\protect\JournalTitle{Proc. SPIE}}  (1989).

\bibitem{conan1a2011integral}
J.-M. Conan, H.-F. Raynaud, C.~Kulcs{\'a}r, S.~Meimon, and G.~Sivo,
  \enquote{Are integral controllers adapted to the new era of {ELT} adaptive
  optics?} {\protect\JournalTitle{Proc. AO4ELT2}}  (2011).

\bibitem{hill}
J.~Hill, R.~Green, D.~Ashby, J.~Brynnel, N.~Cushing, J.~Little, J.~Slagle, and
  R.~Wagner, \enquote{{The Large Binocular Telescope},}
  {\protect\JournalTitle{Proc. SPIE}}  (2012).

\bibitem{agapito2014adaptive}
G.~Agapito, C.~Arcidiacono, F.~Quir{\'o}s-Pacheco, and S.~Esposito,
  \enquote{Adaptive optics at short wavelengths,}
  {\protect\JournalTitle{Experimental Astronomy}} \textbf{37}, 503--523 (2014).

\bibitem{esposito}
S.~Esposito, A.~Riccardi, and B.~Femenía, \enquote{{Differential piston
  angular anisoplanatism for astronomical optical interferometers},}
  {\protect\JournalTitle{Astronomy and Astrophysics}}  (2000).

\end{thebibliography}
\end{document}